\documentclass[aps,showpacs,prb,twocolumn,floatfix]{revtex4}
\usepackage{amsmath,amssymb,graphicx,bm}
\begin{document}
\title{Density and current response functions in strongly disordered
  electron systems: Diffusion, electrical conductivity and Einstein
  relation}

\author{V.  Jani\v{s}} \author{ J. Koloren\v{c}} \author{V. \v{S}pi\v{c}ka}

\affiliation{Institute of Physics, Academy of Sciences of the Czech
  Republic, Na Slovance 2, CZ-18221 Praha 8, Czech Republic}
\email{janis@fzu.cz, kolorenc@fzu.cz, spicka@fzu.cz}

\date{\today}


\begin{abstract}
  We study consequences of gauge invariance and charge conservation of an
  electron gas in a strong random potential perturbed by a weak
  electromagnetic field. We use quantum equations of motion and Ward
  identities for one- and two-particle averaged Green functions to
  establish exact relations between density and current response functions.
  In particular we find precise conditions under which we can extract the
  current-current correlation function from the density-density correlation
  function and vice versa. We use these results in two different ways to
  extend validity of a formula relating the density response function and
  the electrical conductivity from semiclassical equilibrium to quantum
  nonequilibrium systems.  Finally we introduce quantum diffusion via a
  response function relating the averaged current with the negative
  gradient of the averaged charge density. With the aid of this response
  function we derive a quantum version of the Einstein relation and prove
  the existence of the diffusion pole in the zero-temperature electron-hole
  correlation function with the long-range spatial fluctuations controlled
  by the static diffusion constant.
\end{abstract}

\maketitle 

\section{Introduction}
\label{sec:intro}

Low-energy physics of equilibrium systems with weakly interacting electrons
is well understood both qualitatively and quantitatively. The relevant
information about the equilibrium system is contained in one- and
two-particle Green functions, in particular in their behavior around the
Fermi energy. A number of reliable approximate methods have been developed
for the calculation of these functions. Among them, systematic renormalized
perturbation expansions based on the many-body Feynman diagrammatic
technique have proved most effective.  When impurities or quenched
configurational randomness are added we still are able to describe
equilibrium properties of such a system quite reliably within diagrammatic
schemes and a mean-field-type coherent-potential approximation
(CPA).\cite{Elliot74}

The situation gets less straightforward if we come out of equilibrium. This
is the case when the system is disturbed by a time-dependent external force
but does not manage to reach a new equilibrium within the relaxation times
of the experimental setup.  We then have to determine the response of the
system to the external perturbation to obtain experimentally relevant data.
However, we do not have at our disposal many established methods to
calculate nonequilibrium response functions.  When the perturbation is
weak, which is the most common situation in practice, we can use linear
response theory to calculate effectively the response
functions.\cite{Plischke94} A fundamental tool for the calculation of
response functions within the linear-response theory is the Kubo
formalism.\cite{Kubo57} It determines how the response functions of the
disturbed nonequilibrium system can be expressed in terms of two-particle
Green (correlation) functions that are characteristics of an equilibrium
state.  Kubo formulas are means for the description of nonequilibrium
systems with equilibrium functions.  Unlike the response functions, the
equilibrium Green functions obey equations of motion and are suitable for
developing systematic approximations within many-body diagrammatic
techniques.

Kubo formalism, however, provides different representations for different
response functions that we have to approximate separately.  The problem
that emerges with independent approximations for different response
functions via different Kubo formulas is that we are usually unable to keep
exact relations between response functions that may hold due to special
symmetries of the system under investigation. Most pronounced case with
``hidden'' symmetries is the response to a weak electromagnetic
perturbation where we have to satisfy gauge invariance and charge and
current conservations. Instantaneous field-dependent deviations of charge
and current densities from their equilibrium values are described by
density and current response functions. Within linear-response theory these
functions are obtained from independent Kubo formulas with density-density
and current-current correlation functions, respectively. Although both
correlation functions are derived from the same two-particle Green
function, they are normally approximated independently according to the
purpose they serve to.

When quantum coherence is negligible we can calculate the transport
properties from the coherent-potential approximation.\cite{Elliot74} There
is no contribution to the homogeneous current-current correlation function
beyond the single electron-hole bubble in the single-band
coherent-potential approximation.\cite{Velicky69} Hence most theories
beyond the CPA, either on model or realistic level, use the Kubo formula
for the electrical conductivity with the current-current correlation
function to determine transport properties of disordered
solids.\cite{Gonis92,Turek97}

On the other hand, when quantum coherence effects are substantial and we
expect the Anderson metal-insulator transition, electrical conductivity is
usually calculated from the electron-hole correlation function with the aid
of the diffusion constant controlling its low-energy behavior near the
diffusion pole.\cite{Vollhardt92,Belitz94} The diffusion pole is crucial
and enables application of scaling arguments and the renormalization group
approach to Anderson localization.\cite{Lee85}

There is a number of more or less heuristic arguments in the literature
that relate the density response with the conductivity.\cite{Rammer98} They
are essentially based on gauge invariance and charge conservation of a
semiclassical equilibrium description of an electron gas exposed to an
electromagnetic field. In weakly disordered quantum systems (described by
continuum models) a relation between the density response and conductivity
can formally be derived as follows.\cite{Rammer98} Gauge invariance is used
to relate the external scalar potential with the electric field $\mathbf{E}
= -\boldsymbol{\nabla}\varphi$.  The current density generated by the
external field then is
\begin{equation}
  \label{eq:induced_current}
  \mathbf{j}({\bf q},\omega) = \boldsymbol{\sigma}({\bf q},\omega)\cdot
  \mathbf{E} ({\bf q},\omega) = -  i \boldsymbol{\sigma}({\bf
q},\omega)\cdot   {\bf q}\ \varphi({\bf q},\omega)
\end{equation}
where $\boldsymbol{\sigma}({\bf q},\omega)$ denotes tensor of the
electrical conductivity.  Charge conservation is expressed by a continuity
equation. In equilibrium we can use its operator form that follows from the
Heisenberg equations of motion for the current and density operators. For
Hamiltonians with quadratic dispersion relation we have
\begin{equation}
  \label{eq:continuity_equation}
   e \partial_t\widehat{n}(\mathbf{x},t) + \boldsymbol{\nabla}\cdot
  \widehat{\mathbf{j}}(\mathbf{x},t) = 0 \ .
\end{equation}
Energy-momentum representation of the continuity equation in the
ground-state solution is
\begin{equation}
  \label{eq:continuity_Fourier}
  - i\omega e\delta n({\bf q},\omega) +i \mathbf{q}\cdot\mathbf{j}({\bf
  q},\omega) = 0 \ .
\end{equation}
We have to use a density variation of the equilibrium density, i.~e.,
the externally induced density $\delta n({\bf q},\omega) = n({\bf
  q},\omega) -n_0$ in the continuity equation with averaged values of
operators.  From the above equations and for linear response $\delta
n({\bf q},\omega) = -e\chi({\bf q},\omega) \varphi({\bf q},\omega)$ we
obtain in the isotropic case
\begin{equation}
  \label{eq:Einstein_general}
  \sigma({\bf q},\omega) = \frac{-ie^2\omega}{q^2} \chi({\bf q},\omega) \ .
\end{equation}
The derived equality formally holds for complex functions without
restrictions on momenta or frequencies. Frequencies can, in principle,
be even complex. Relation \eqref{eq:Einstein_general} is often taken
as granted for the whole range of the disorder strength and used for
the definition of the zero-temperature conductivity when describing
Anderson localization transition.\cite{Vollhardt92,Belitz94}

Although the above derivation may seem very general it suffers from a few
flaws. First, the operator continuity equation
\eqref{eq:continuity_equation} cannot be directly used out of equilibrium.
A new term due to causality of the response functions enters the continuity
equation when the equilibrium is disturbed.  Moreover the nonequilibrium
density and current operators in linear-response theory no longer obey
Heisenberg equations of motion with the perturbed Hamiltonian.  The
perturbation is decoupled from the equilibrium Hamiltonian and is treated
only to linear order. Second, strongly disordered electron systems can be
desribed only by lattice models with a nonquadratic dispersion relation.
Continuity equation \eqref{eq:continuity_equation} is then to be modified
even in equilibrium.  Hence the above, in the literature broadly
disseminated reasoning leading to Eq.~\eqref{eq:Einstein_general} cannot be
fully trusted in quantitative studies of strongly disordered systems out of
equilibrium.

Presently there is no reliable theory of strongly disordered electrons
beyond the mean-field CPA. Since mean-field approximations do not include
vertex corrections to the electrical conductivity they are not suitable for
the investigation of localization effects in three spatial dimensions.
Unlike low dimensions ($d\le2$), Anderson localization in $d=3$ may occur
only in strongly disordered systems. Anderson localization in bulk systems
has not yet been understood or quantitatively described in a satisfactory
manner.  Its quantitative description demands bridging the gap between the
mean-field (CPA) transport theory and theories for weakly and strongly
localized electron states. It is clear that such an interpolating scheme
should be based on advanced approximations for two-particle functions.
 
Recently a diagrammatic method for summations of classes of two-particle
diagrams has been proposed.\cite{Janis01b} This theory has a potential to
interpolate between the mean-field and localization theories provided exact
relations between the density and current response functions have been
proved for this scheme in the whole range of the disorder strength.
Criteria for validity of relations between density-density and
current-current correlation functions in approximate treatments of strongly
disordered systems have not yet been established. When we want to go beyond
the CPA we still have to establish such relations in the metallic regime of
tight-binding models with extended electron states.  Simultaneously we have
to formulate conditions under which the derived relations hold or may be
broken.

The aim of this paper is threefold. First, we derive various
relations between current-current, current-density, and density-density
correlation functions for strongly disordered electrons out of equilibrium,
i.~e., beyond the reach of the existing derivations. Second, we articulate
conditions under which these relations hold or to which extent they may be
broken in quantitative treatments and show when the conductivity can be
calculated from the density response function and vice versa. Third, we
introduce a generalization of diffusion via a quantum response function,
test validity of the Einstein relation for this quantum diffusion, and
relate quantum diffusion with its classical counterpart and with the
diffusion pole in the electron-hole correlation function.

To reach this goal we use Kubo formalism and averaged many-body Green
functions.  In this approach a weaker form of the continuity equation
expressed in terms of one- and two-particle Green functions replaces the
operator identity. It is derived from equations of motion for Green
functions and Ward identities between them. We use two Ward identities that
are not fully equivalent. One expresses conservation of probability and the
other reflects charge conservation.  Each Ward identity is used in a
different manner to relate the conductivity with the density response.

The layout of the paper is as follows. In Sec.~\ref{sec:density_current} we
summarize the definitions and useful representations of the density and
current response functions. In Sec.~\ref{sec:Einstein_relation} we show how
the current-current correlation function emerges from a momentum expansion
of the density response function. Then in
Sec.~\ref{sec:dynamical_conductivity} we derive continuity equations
expressed in Green functions and use them to show how the density response
function can be revealed from the conductivity. Based on the exactly
derived relation between the density and current response function we
introduce in Sec.~\ref{sec:Diff_Einstein} quantum diffusion, derive the
diffusion pole in the hydrodynamic limit of the density response function
and relate the diffusion constant to the conductivity. In
Sec.~\ref{sec:infinite_d} we illustrate general results on an exactly
solvable limit of infinite spatial dimensions. In Appendix we discuss
assumptions for and the range of validity of the Vollhardt-W\"olfle-Ward
identity used in the derivation of the continuity equation for Green
functions.\\

\section{Density and current response functions in disordered systems}
\label{sec:density_current}

A simplest description of the electron motion in impure, weakly correlated
metals is provided by a tight-binding Anderson model.  It assumes
noninteracting spinless electrons moving in a random, site-diagonal
potential $V_i$. Its Hamiltonian reads
\begin{eqnarray}\label{eq:AD_hamiltonian}
  \widehat{H}_{AD}&=&\sum_{<ij>}t_{ij}c_{i}^{\dagger} c_{j}+\sum_iV_ic_{i
    }^{\dagger } c_{i} \, .
\end{eqnarray}
The values of the random potential $V_i$ are site-independent and obey a
disorder distribution $\rho(V)$. I.~e., a function depending on the random
potential $V_i$ is averaged as
\begin{eqnarray}   \label{eq:averaging}
  \left\langle X(V_i)\right\rangle_{av}&=&\int_{-\infty}^{\infty}
  dV_i\rho(V_i)X(V_i)\, .
\end{eqnarray}
The averaged two-particle propagator (resolvent) is defined as the averaged
product of one-particle propagators
\begin{multline}
  \label{eq:av_2PP}
  G^{(2)}_{ij,kl}(z_1,z_2)= \\ \left\langle\left[z_1\widehat{1}-\widehat{t}
      -\widehat{V}\right]^{-1}_{ij} \left[z_2\widehat{1}-\widehat{t}
      -\widehat{V}\right]^{-1}_{kl} \right\rangle_{av} \ .
\end{multline}
Fourier transform to momentum space is not uniquely defined, since due to
momentum conservation we have only three independent momenta.  For our
purposes we choose the following notation and definition of the Fourier
transform
\begin{multline}\label{eq:2P_momentum}
  G^{(2)}_{{\bf k}{\bf k}'}(z_1,z_2;{\bf q})= \frac
  1{N}\sum_{ijkl}e^{-i({\bf k} + {\bf q}/2){\bf R}_i} e^{i({\bf k}'+{\bf
      q}/2){\bf R}_j}\\ \times e^{-i({\bf k}'-{\bf q}/2){\bf R}_k}
  e^{i({\bf k} - {\bf q}/2){\bf R}_l} G^{(2)}_{ij,kl}(z_1,z_2)\, .
\end{multline}

The density response function is determined from the following
thermodynamic representation of the Kubo formula\cite{Enz92}
\begin{multline} \label{eq:DRF_thermo}
  \chi({\bf q},i\nu_m) =\\ -\frac 1{N^2}\sum_{{\bf k}{\bf k}'} k_BT \!\!
  \sum_{n=-\infty}^{\infty} G^{(2)}_{{\bf k}{\bf k}'}(i\omega_n,i\omega_n +
  i\nu_m;{\bf q}) \ .
\end{multline}
Here $\omega_n=(2n+1)\pi T$ are fermionic and $\nu_m=2m\pi T$ bosonic
Matsubara frequencies at temperature $T$. Due to the analytic properties of
the two-particle Green function we can perform analytic continuation from
imaginary Matsubara to real frequencies using contour integrals. Only the
contribution along the real axis remains and we obtain
\begin{widetext}
\begin{multline}
  \label{eq:DRF_real}
  \chi({\bf q},\omega+i0^+) = \frac 1{N^2}\sum_{{\bf k}{\bf k}'}
  \int_{-\infty}^{\infty} \frac {d E}{2\pi i} \left\{ \left[f(E+\omega) -
      f(E)\right]
    G^{AR}_{{\bf k}{\bf k}'}(E,E+\omega;{\bf q}) \right. \\[2pt]
  \left. + \ f(E) G^{RR}_{{\bf k}{\bf k}'}(E,E+\omega;{\bf q}) -
    f(E+\omega) G^{AA}_{{\bf k}{\bf k}'}(E,E+\omega;{\bf q})\right\} \ .
\end{multline}
\end{widetext}
We used the following definition
\begin{equation*}
  G^{AR}_{{\bf k}{\bf k}'}(E,E+\omega;{\bf q}) = G^{(2)}_{{\bf k}{\bf
      k}'}(E - i0^+,E + \omega + i0^+;{\bf q})
\end{equation*}
and analogously for the functions $G^{RR}$ and $G^{AA}$, where both
energies are from the upper and lower complex energy half-plane,
respectively. We denoted the Fermi function%
$f(E)=[1+\exp\{\beta (E-\mu)\}]^{-1}$ with the chemical potential $\mu$.

The two-particle Green function is generally determined from a
Bethe-Salpeter equation in which the input is a two-particle irreducible
vertex. The latter is known only approximately, except for special limits.
However, a few specific elements of the full two-particle Green function
can be evaluated without knowing the two-particle irreducible vertex. It is
sufficient to know only the one-particle Green function to determine them.
Determination of two-particle functions from one-particle ones is enabled
by Ward identities. The Ward identity relating the averaged one- and
two-particle Green functions reads
\begin{multline}
  \label{eq:VW_momentum}
  \frac 1N\sum_{{\bf k}'}G^{(2)}_{{\bf k}{\bf k}'}(z_1,z_2;{\bf 0})= \frac
  1{z_2 - z_1} \left[ G({\bf k},z_1)\right. \\ \left. -\ G({\bf
      k},z_2)\right]\, .
\end{multline}
and was proved for the first time within the coherent-potential
approximation by Velick\'y.\cite{Velicky69} It is a nonperturbative
identity valid quite generally beyond the CPA.\cite{Janis01b} As a
consequence of this identity we obtain vanishing of the density response
function for a homogeneous global perturbation, i.~e., for ${\bf q}=0$.
Using \eqref{eq:VW_momentum} in \eqref{eq:DRF_real} we easily find
\begin{multline}\label{eq:DRF_homo}
  \chi({\bf 0},\omega + i0^+)\\ = \frac 1\omega \int_{-\infty}^{\infty}
  \frac
  {d E}{2\pi i} \left\{ f(E+\omega)\left[G^A(E+\omega) \right.\right. \\
  \left.\left. -\ G^R(E+\omega)\right] - f(E) \left[G^A(E) -
      G^R(E)\right]\right\} = 0\ .
\end{multline}

Another situation where we do not need to know the two-particle irreducible
vertex is the static limit, $\omega=0$. First, the static density response
function is real and reads
\begin{subequations}\label{eq:DRF_T}
\begin{equation}\label{eq:DRF_static}
  \chi({\bf q},0) = \frac 1{N^2}\sum_{{\bf k}{\bf k}'}
  \int_{-\infty}^{\infty} \frac {d E}{\pi} f(E) \Im G^{RR}_{{\bf k}{\bf
      k}'}(E,E;{\bf q})
\end{equation}
that in the limit ${\bf q}\to 0$ goes over to
\begin{equation}\label{eq:DRF_static_zero}
    \chi({\bf 0},0) = \int_{-\infty}^\infty \frac{d\omega}{\pi}\
    \frac{\partial f(\omega)}{\partial\omega}\ \Im G^R(\omega)\
    \xrightarrow[T\to0]{}  n_F \
\end{equation}
\end{subequations}
with $n_F$ as the density of states at the Fermi level and
$G^R(\omega)=N^{-1}\sum_\mathbf{q} G^R(\mathbf{q},\omega)$. Note that
Eqs.~\eqref{eq:DRF_T} hold only in the limit $\omega/q\to0$.  In the
inverse case, $q/\omega\to0$, Eq.~\eqref{eq:DRF_homo} applies.
Noncommutativity of the limits $\omega\to0$ and $q \to0$ indicates that the
point $q=0, \omega=0$ is not analytic in the same manner as in the Fermi
liquid theory.\cite{Rickayzen80}

The current-current correlation function $\Pi_{\alpha\beta}({\bf q},t)$
relates the current of the system $\mathbf{j}(\mathbf{q},t)$ perturbed by
an external vector potential $\mathbf{A}(\mathbf{q},t)$.  Tensor of the
electrical conductivity $\sigma_{\alpha\beta}({\bf q},t)$ determines the
current response to an external electric field. It can be obtained from the
current response function by using a relation
$\mathbf{E}(\mathbf{x},t)=\dot{\mathbf{A}}(\mathbf{x},t)$.  In the Fourier
representation we obtain\cite{Mahan00}
\begin{equation}\label{eq:conductivity}
  \sigma_{\alpha\beta}({\bf q},\omega_+) = \frac
  i{\omega}\left[\Pi_{\alpha\beta}({\bf q},\omega_+) -
   \Pi_{\alpha\beta}({\bf q},0)\right]
\end{equation}
with $\omega_+=\omega+i0^+$. The compensation term $\Pi_{\alpha\beta}({\bf
  q},0)$ on the right-hand side of Eq.~\eqref{eq:conductivity} is real and
was introduced to warrant finiteness of the complex conductivity. This
additive term is not generated from the Kubo formula with a commutator of
current operators but originates from gauge invariance of the electron gas
in an external electromagnetic field.\cite{Note1}
According to Eq.~\eqref{eq:conductivity}  there is no current in the system
with a static vector potential and the static (complex) conductivity
can be defined only via a dynamical one from the limit $\omega\to0$.

Unlike the density response function the current response function
cannot be simplified in any limit and to determine the conductivity
from Eq.~\eqref{eq:conductivity} we have to know the two-particle
irreducible vertex. The current response function alike the density
response function is a complex quantity with the same analytic
properties.  Of interest for us are the parallel components that for
real frequencies read
\begin{widetext}
\begin{multline}\label{eq:T_conductivity}
  \sigma_{\alpha\alpha}({\bf q},\omega)= -\frac{e^2}{4}\frac
  1{N^2}\sum_{{\bf k},{\bf k}'} \left[v_\alpha({\bf
      k}^{\phantom\prime}+{\bf q}/2) - v_\alpha({-\bf k}+{\bf
      q}/2)\right]\left[v_\alpha({\bf k}'+{\bf q}/2) - v_\alpha({-\bf
      k}'+{\bf q}/2)\right]\\ \times \int_{-\infty}^{\infty}\frac {d E}{2
    \pi\omega} \left\{[f(E+\omega) - f(E)] G^{AR}_{{\bf k}{\bf
        k}'}(E,E+\omega;{\bf q}) + f(E) G^{RR}_{{\bf k}{\bf
        k}'}(E,E+\omega;{\bf q}))\right. \\ \left.  -
    f(E+\omega)G^{AA}_{{\bf k}{\bf k}'}(E,E+\omega;{\bf q}) - f(E)[
    G^{RR}_{{\bf k}{\bf k}'}(E,E;{\bf q})- G^{AA}_{{\bf k}{\bf
        k}'}(E,E;{\bf q})] \right\}\ .
\end{multline}
\end{widetext}
We denoted the group velocity $v_\alpha({\bf k})=m^{-1}\partial/\partial
k_\alpha \epsilon({\bf k})$, where $\epsilon({\bf k})$ is the dispersion
relation of the underlying lattice.

\section{Dynamical conductivity calculated from the  density response}
\label{sec:Einstein_relation}

As first we show how the dynamical conductivity can be obtained from the
long-range behavior of the density response function (hydrodynamic limit of
small transfer momentum). To this purpose we use the Velick\'y-Ward
identity, Eq.~\eqref{eq:VW_momentum}. It enables us to evaluate averaged
matrix elements of the two-particle Green function with zero transfer
momentum in terms of the one-particle propagator.  Zero transfer momentum
is a severe restriction on the applicability and utilization of this Ward
identity. Physically we are interested in the hydrodynamic limit, i.~e.,
the asymptotics $q\to0$. We are unable to extend identity
\eqref{eq:VW_momentum} beyond the homogeneous case.  However, if we assume
analyticity of the asymptotics $q\to0$, we can use momentum $q$ as an
expansion parameter and investigate the hydrodynamic limit perturbatively.
The hydrodynamic limit is analytic if frequency $\omega\neq0$, i.~e., we
are in the regime $q/\omega\ll1$. We then can expand the density response
function in powers of momentum $q$.

We define two configuration-dependent resolvent operators
\begin{equation}
  \label{eq:resolvent}
  \widehat{G}_\pm(z) = \left[z_\pm\widehat{1} - \widehat{t} \mp
  \Delta_1\widehat{t} - \Delta_2 \widehat{t} - \widehat{V} \right]^{-1}
\end{equation}
where $\Delta_1 \widehat{t} \pm \Delta_2\widehat{t}$ is a difference in
the dispersion relation of the two resolvents and is defined via its matrix
elements
$\langle{\bf k}|\Delta_1 \widehat{t} |{\bf k}'\rangle = \delta(\mathbf{k}'
-\mathbf{k})\;\mathbf{v}(\mathbf{k})\cdot\mathbf{q}/2$, $\langle {\bf%
  k}|\Delta_2 \widehat{t} |{\bf k}'\rangle = \delta(\mathbf{k}' -%
\mathbf{k})\;(\mathbf{q}\cdot\nabla_k)^2\epsilon(\mathbf{k})/8$. We will be
interested in the following function of two energies
\begin{equation}
  \label{eq:trace_2resolvent}
  I(z_+,z_-;\mathbf{q})  = \frac 1N\mbox{Tr}\left[\widehat{G}_+(z_+)
\widehat{G}_-(z_-)   \right]
\end{equation}
that can be, after averaging over the configurations of the random
potential, expressed via the two-particle Green function
\begin{equation}\label{eq:trace_2resolvent_av}
   \langle I(z_+,z_-;\mathbf{q})\rangle_{av} = \frac
1{N^2}\sum_{\mathbf{k}\mathbf{k}'}
G^{(2)}_{\mathbf{k}\mathbf{k}'}(z_+,z_-;\mathbf{q})\ .
\end{equation}

We assume that the expansion in the operators $\Delta_{1,2} \widehat{t}$
commutes with the configurational averaging. We first expand the quantity
$I$ and then average the series term by term. It is sufficient for our
purposes to expand only to second order in momentum $\mathbf{q}$. This
precision determines the leading small-momentum behavior.

Expanding quantity $I$ in $\Delta_{1,2} \widehat{t}$ we have to keep the
order of operators in the products, since $\Delta_{1,2} \widehat{t}$ and
the resolvent $\widehat{G}(z)$ do not commute. The expansion to second
order reads
\begin{widetext}
\begin{multline} \label{eq:trace_expansion}
I_2(z_+,z_-\mathbf{q}) = \mbox{Tr}\left\{\widehat{G}(z_+)\widehat{G}(z_-) +
\widehat{G}(z_+) \left[\Delta_{1} \widehat{t} + \Delta_{2}
\widehat{t}\right] \widehat{G}(z_+) \widehat{G}(z_-) \right. \\ \left. -\
\widehat{G}(z_+)\widehat{G}(z_-) \left[\Delta_{1} \widehat{t} - \Delta_{2}
\widehat{t}\right] \widehat{G}(z_-) - \widehat{G}(z_+) \Delta_{1}
\widehat{t} \widehat{G}(z_+) \widehat{G}(z_-) \Delta_{1} \widehat{t}
\widehat{G}(z_-)\right. \\ \left. +\ \widehat{G}(z_+)\left[\Delta_{1}
\widehat{t} \widehat{G}(z_+) \Delta_{1} \widehat{t}\widehat{G}(z_+) +
\widehat{G}(z_-) \Delta_{1} \widehat{t} \widehat{G}(z_-) \widehat{\Delta_1
t} \right] \widehat{G}(z_-)\right\} \ .
\end{multline}
Each direct product of the resolvent operators $\widehat{G}(z_\pm)
\widehat{G}(z_\mp)$ can be simplified using identity
\eqref{eq:VW_momentum}. Doing this consequently we end up with a sum of
products of two resolvents. We have three different terms to analyze:
\begin{subequations}\label{eq:difference}
\begin{multline} \label{eq:difference1}
  \mbox{Tr}\left\{\widehat{G}(z_+)\left[ \left( \Delta_{1} \widehat{t} +
        \Delta_{2} \widehat{t}\right) \widehat{G}(z_-) - \widehat{G}(z_+)
      \left( \Delta_{1} \widehat{t} - \Delta_{2} \widehat{t}\right)
    \right]\widehat{G}(z_-)\right\} = \\ \frac 1{z_+ - z_-} \left\{
    \mbox{Tr}\left[\left(\widehat{G}'(z_+) + \widehat{G}'(z_-)\right)
      \Delta_{1} \widehat{t}\right] +\mbox{Tr}\left[\left(\widehat{G}'(z_+)
        -
        \widehat{G}'(z_-)\right) \Delta_{2} \widehat{t}\right]\right. \\
  \left.  -\ \frac 2 {z_+ - z_-}\mbox{Tr}\left[\Delta_{1} \widehat{t}
      \left(\widehat{G}(z_+)- \widehat{G}(z_-)\right)\right]\right\}\ ,
\end{multline}
\begin{multline} \label{eq:difference2}
  \mbox{Tr}\left\{\widehat{G}(z_+)\Delta_{1} \widehat{t} \widehat{G}(z_+)
    \widehat{G}(z_-)\Delta_{1} \widehat{t} \widehat{G}(z_-)\right\} = \frac
  1{(z_+ - z_-)^2}\left\{\mbox{Tr}\left[\widehat{G}(z_+) \Delta_{1}
      \widehat{
        t}\widehat{G}(z_+) \Delta_{1} \widehat{t}\right]\right. \\
  \left. +\ \mbox{Tr}\left[\widehat{G}(z_-) \widehat{\Delta_1
        t}\widehat{G}(z_-) \Delta_{1} \widehat{t}\right] -
    2\mbox{Tr}\left[\widehat{G}(z_+)\Delta_{1} \widehat{t}
      \widehat{G}(z_-)\Delta_{1} \widehat{t}\right] \right\}\ ,
\end{multline}
\begin{multline} \label{eq:difference3}
  \mbox{Tr}\left\{\widehat{G}(z_+)\left[\Delta_{1} \widehat{t}
      \widehat{G}(z_+) \Delta_{1} \widehat{t}\widehat{G}(z_+) +
      \widehat{G}(z_-) \Delta_{1} \widehat{t} \widehat{G}(z_-) \Delta_{1}
      \widehat{t} \right]
    \widehat{G}(z_-)\right\} =\\
  \frac 1{z_+ - z_-}\left\{\mbox{Tr}\left[\widehat{G}'(z_+) \Delta_{1}
      \widehat{
        t}\widehat{G}(z_+) \Delta_{1} \widehat{t}\right]
    - \mbox{Tr}\left[\widehat{G}'(z_-) \Delta_{1}
      \widehat{t}\widehat{G}(z_-) \Delta_{1} \widehat{t}\right] \right. \\
  \left.  -\ \frac 1{z_+ - z_-}\mbox{Tr}\left[\left(\widehat{G}(z_+) -
        \widehat{G}(z_-)\right) \Delta_{1}
      \widehat{t}\left(\widehat{G}(z_+) - \widehat{G}(z_-)\right)
      \Delta_{1} \widehat{t}\right]\right\} \ .
\end{multline}
\end{subequations}
\end{widetext}

We insert Eqs.~\eqref{eq:difference} in expansion
\eqref{eq:trace_expansion} and average over the configurations of the
random potential $V_i$. To recover the density response function we have
to limit the complex energies to the real axis from above or below with
which we distinguish the causality. We define correlation functions that
can be represented as a trace of the averaged two-particle Green function
\begin{equation}
  \label{eq:Phi}
   \Phi^{ab}_E({\bf q}, \omega) = \frac 1{N^2} \sum_{{\bf k}{\bf
  k}'}G^{ab}_{{\bf k}{\bf k}'} (E,E + \omega;{\bf q})\ .
\end{equation}
where $a,b$ stand for $A,R$ depending on whether the imaginary part of the
corresponding frequency argument is positive or negative.
We then can represent the leading asymptotics in small momenta ${\bf q}$ of
the averaged electron-hole correlation function
\begin{widetext}
\begin{subequations}\label{eq:G1}
\begin{multline}  \label{eq:GAR1}
  \Phi^{AR}_E({\bf q}, \omega)- \Phi^{AR}_E({\bf 0}, \omega)= -\frac
  1{2\omega}\left[\left\langle \partial_E \left(G^R(E+\omega)
        +G^A(E)\right) \mathbf{q}\cdot\mathbf{v}\right\rangle\right]\\
  +\frac 1{8\omega}\left[\left\langle \partial_E \left(G^R(E+\omega)
        -G^A(E)\right) (\mathbf{q}\cdot\nabla)^2\epsilon
    \right\rangle\right] + \frac 1{\omega^2}
  \left[\left\langle\left(G^R(E+\omega) - G^A(E)\right)
      \mathbf{q}\cdot\mathbf{v}\right\rangle\right] \\
  - \frac 1{4\omega N^2} \sum_{{\bf k}{\bf
      k}'}\mathbf{q}\cdot\mathbf{v}({\bf k})\;
  \mathbf{q}\cdot\mathbf{v}({\bf k}')\left\{\partial_\varepsilon
    \left[G^{AA}_{{\bf k}{\bf k}'}(E+\varepsilon, E;{\bf 0}) - G^{RR}_{{\bf
          k}{\bf k}'}(E+\omega, E+\omega + \varepsilon;{\bf
        0})\right]_{\varepsilon=0} \right. \\ \left.  +\ \frac
    2{\omega}\left[G^{AA}_{{\bf k}{\bf k}'}(E, E;{\bf 0}) + G^{RR}_{{\bf
          k}{\bf k}'}(E+\omega, E+\omega;{\bf 0}) - 2 G^{AR}_{{\bf k}{\bf
          k}'}(E, E+\omega;{\bf 0})\right] \right\}
\end{multline}
and analogously of the electron-electron ($\Phi^{RR}$) and hole-hole
($\Phi^{AA}$) ones that can be written generically as
\begin{multline}  \label{eq:GRR1}
  \Phi^{aa}_E({\bf q}, \omega)- \Phi^{aa}_E({\bf 0}, \omega) = -\frac
  1{2\omega}\left[\left\langle \partial_E \left(G^{a}(E+\omega) +
        G^{a}(E)\right) \mathbf{q}\cdot\mathbf{v}\right\rangle\right]\\
  +\frac{1}{8\omega}\left[\left\langle \partial_E \left(G^a(E+\omega)
        -G^a(E)\right) (\mathbf{q}\cdot\nabla)^2\epsilon
    \right\rangle\right] + \frac 1{\omega^2}
  \left[\left\langle\left(G^{a}(E+\omega) - G^{a}(E)\right)
      \mathbf{q}\cdot\mathbf{v}\right\rangle\right] \\ - \frac 1{4\omega
    N^2} \sum_{{\bf k}{\bf k}'}\mathbf{q}\cdot\mathbf{v}({\bf k})\;
  \mathbf{q}\cdot \mathbf{v}({\bf k}')\ \left\{\partial_\varepsilon
    \left[G^{aa}_{{\bf k}{\bf k}'}(E+\varepsilon, E;{\bf 0}) - G^{aa}_{{\bf
          k}{\bf k}'}(E+\omega, E+\omega + \varepsilon;{\bf
        0})\right]_{\varepsilon=0} \right.\\ \left.  +\ \frac
    2{\omega}\left[G^{aa}_{{\bf k}{\bf k}'}(E, E;{\bf 0}) + G^{aa}_{{\bf
          k}{\bf k}'}(E+\omega, E+\omega;{\bf 0}) - 2 G^{aa}_{{\bf k}{\bf
          k}'}(E, E+\omega;{\bf 0})\right] \right\} \ .
\end{multline}
\end{subequations}
\end{widetext}
In the above equations we used angular brackets to denote summation over
fermionic momenta in the first Brillouin zone of one-electron functions,
that is
\begin{equation}
  \label{eq:1P_summed}
  \left\langle G(\omega)\ f_{\bf q} \right\rangle = \frac 1N \sum_{\bf k}
G({\bf   k},\omega)\ f_{\bf q}(\mathbf{k})\ .
\end{equation}
Note that only the term with $\omega^{-2}$ from the electron-hole
correlation function diverges in the limit $\omega\to0$ while in the
electron-electron (hole-hole) functions it remains finite even in the
static limit. This is a manifestation of the diffusion pole missing the
latter two correlation functions.

In Eq.~\eqref{eq:G1} different powers of momentum $\mathbf{q}$ and
frequency $\omega$ appear. There is no restriction on validity of
Eqs.\eqref{eq:G1} in frequency but it holds only for small momenta, more
precisely only perturbatively up to second order.  Equations \eqref{eq:G1}
establish relations between the density-density and the current-current
correlation functions in the asymptotic limit $q=0$.  Generally, however,
these two functions are not directly proportional, since the right-hand
sides of Eqs.\eqref{eq:G1} contain one-particle contributions.  They cancel
each other if we combine the electron-hole with electron-electron and
hole-hole correlation functions appropriately to build up the density
response function.

For isotropic situations we define in the hydrodynamic limit
\begin{subequations}\label{eq:mobility}
\begin{equation}
  \label{eq:mobility_iso}
  g(\omega) =  -i \lim_{q\to 0} \frac \omega{q^2} \chi({\bf q},\omega) \ .
\end{equation}
This quantity can be generalized to anisotropic cases as
\begin{equation}
  \label{eq:mobility_non}
  g_{\alpha\beta}(\omega) = -i \omega \ \frac{\partial^2}{\partial q_\alpha
    \partial q_\beta} \chi({\bf q},\omega)\big|_{q=0} \ .
\end{equation}
\end{subequations}
Function $g(\omega)$ measures leading long-range correlations of the
density response function. Its real part can be identified with the
diffusive conductivity or mobility of the system.  It is now easy to find
an explicit representation of this function using Eqs.~\eqref{eq:G1}.  All
contributions except for the last terms on the right-hand sides of
Eqs.~\eqref{eq:G1} cancel and we obtain equality $\sigma_{\alpha\alpha}({\bf
  0},\omega) = e^2g_{\alpha\alpha}(\omega)$ being just
Eq.~\eqref{eq:Einstein_general} in the limit ${\bf q}={\bf 0}$.  Note that
in the course of the derivation we had to sum contributions from the
electron-hole, electron-electron, and hole-hole correlation functions and
integrate over energies.

If we resort to the low-frequency limit $\omega\to0$, the diffusion pole in
the electron-hole correlation function dominates in the density response
and we recover the conductivity directly from $\Phi^{AR}$. It is
interesting to note that in this static limit we derive the mobility
($dc$-conductivity) with contributions from $G^{AR}$ and $G^{RR}$ solely
from the electron-hole correlation function $\Phi^{AR}$.

\section{Density response function calculated from the conductivity}
\label{sec:dynamical_conductivity}

We showed in the preceding section how the conductivity can be revealed
from the small-momentum behavior of the density response function.  In this
section we express the current-current correlation function in terms of the
density-density correlation function. The aid we use for this task is a
continuity equation and another Ward identity
\begin{multline}
  \label{eq:VWW_identity}
  \Sigma({\bf k}_+,z_+) - \Sigma({\bf k}_-, z_-) = \frac 1N \sum_{{\bf
      k}'}\Lambda_{{\bf k}{\bf k}'}(z_+,z_-;{\bf q})\\ \times \left[G({\bf
      k}'_+,z_+) - G({\bf k}'_-, z_-) \right] \ .
\end{multline}
proved for retarded and advanced functions by Vollhardt and W\"olfle in
Ref.~\onlinecite{Vollhardt80b}. We denoted ${\bf k}_\pm={\bf k}\pm{\bf
  q}/2$.  Note that identity \eqref{eq:VWW_identity} for
$\mathbf{q}=\mathbf{0}, z_+=E_F+i0^+,z_-=E_F-i0^+$ was already used in
Ref.~\onlinecite{Maleev75}.  Identity \eqref{eq:VWW_identity} is related to
Eq.~\eqref{eq:VW_momentum}, however, the two Ward identities are identical
neither in the derivation nor in the applicability and validity domains.
The latter holds for nonzero transfer momentum $\mathbf{q}$, i.~e., for an
inhomogeneous perturbation while the former only for $q=0$.  On the other
hand, we show in the appendix that Eq.~\eqref{eq:VWW_identity}, unlike
Eq.~\eqref{eq:VW_momentum}, holds only perturbatively within an expansion in
powers of the random potential. I.~e., we can prove this identity for
nonzero transfer momentum $\mathbf{q}$ only by assuming that perturbation
expansions for the self-energy $\Sigma$ and simultaneously for the
two-particle irreducible vertex $\Lambda$ converge.  It was shown earlier
that in the homogeneous case, $q=0$, the Vollhardt-W\"olfle-Ward identity
follows from the Velick\'y-Ward one.\cite{Janis01b}

To derive a continuity equation for Green functions we start with an
equation of motion for the two-particle Green function. It is a
Bethe-Salpeter equation where the input is a two-particle irreducible
vertex. We have three possibilities (topologically distinct scattering
channels) how to construct a Bethe-Salpeter equation.\cite{Janis01b} For
our purposes the electron-hole channel is the relevant or most suitable
one. There the Bethe-Salpeter equation in momentum representation reads
\begin{widetext}
\begin{equation}
  \label{eq:BS_equation}
   G^{(2)}_{{\bf k}{\bf k}'}(z_+,z_-;{\bf q}) = G({\bf k}_+,z_+)
   G({\bf k}_-, z_-)\left[\delta(\mathbf{k} - \mathbf{k}') + \frac 1N
   \sum_{{\bf k}''}\Lambda_{{\bf k}{\bf k}''}(z_+,z_-;{\bf q})
   G^{(2)}_{{\bf k}''{\bf k}'}(z_+,z_-;{\bf q}) \right]
\end{equation}
\end{widetext}
where $\Lambda_{{\bf k}{\bf k}'}(z_+,z_-;{\bf q})$ is the two-particle
irreducible vertex from the electron-hole channel. One-electron propagators
$G({\bf k}_\pm, z_\pm)=[z_\pm -\Sigma(\mathbf {k}\pm\mathbf{q}/2, z_\pm)
-\epsilon(\mathbf{k} \pm \mathbf{q}/2)]^{-1}$ are the averaged resolvents.

The product of the one-electron propagators from the right-hand side of
Eq.~\eqref{eq:BS_equation} can be decomposed into
\begin{multline}
  \label{eq:2P_product}
  G({\bf k}_+,z_+)G({\bf k}_-, z_-) =\\ - \frac{\Delta_{\bf q} G({\bf
      k};z_+,z_-)}{\Delta z - \Delta_{\bf q}\Sigma({\bf k};z_+,z_-)-
    \Delta_{\bf q}\epsilon({\bf k})}
\end{multline}
where we denoted
\begin{subequations}
\begin{equation}
  \label{eq:energy_variation}
  \Delta_{\bf q} \epsilon({\bf k}) = \epsilon({\bf k}_+) -
  \epsilon({\bf k}_-) \ ,
\end{equation}
\begin{equation}
  \label{eq:Delta_G}
  \Delta_{\bf q} G ({\bf k};z_+,z_-) = G({\bf k}_+,z_+) -  G({\bf
    k}_-,z_-)\ .
\end{equation}
and analogously the difference $\Delta_{\bf q}\Sigma({\bf k};z_+,z_-)$.
\end{subequations}
We multiply both sides of the Bethe-Salpeter equation
\eqref{eq:BS_equation} by the denominator from the right-hand side of
Eq.~\eqref{eq:2P_product} and assume validity of
Eq.~\eqref{eq:VWW_identity} relating the two-particle irreducible vertex
$\Lambda$ with the one-particle irreducible self-energy $\Sigma$. We then
obtain a "difference" equation of motion
\begin{widetext}\begin{multline}
   \label{eq:difference_BSE}
   \left[\Delta_\mathbf{q}\epsilon({\bf k}) - \Delta z\right] G^{(2)}_{{\bf
       k}{\bf k}'}(z_+,z_-;{\bf q}) = \Delta_\mathbf{q} G({\bf
     k};z_+,z_-)\delta(\mathbf{k} - \mathbf{k}') + \frac 1N \sum_{{\bf
       k}''}\Lambda_{{\bf k}{\bf k}''}(z_+,z_-;{\bf q}) \\
   \times \left[\Delta_\mathbf{q} G({\bf k};z_+,z_-)G^{(2)}_{{\bf k}''{\bf
         k}'}(z_+,z_-;{\bf q}) - G^{(2)}_{{\bf k}{\bf k}'}(z_+,z_-;{\bf q})
     \Delta_\mathbf{q} G({\bf k}'';z_+,z_-)\right]\ .
\end{multline}\end{widetext}
We now define correlation functions generalized to complex frequencies
describing density-density and density-current correlations by summing over
the fermionic momenta in Eq.~\eqref{eq:difference_BSE}
\begin{subequations}\label{eq:Phi_def}
\begin{align}
  \label{eq:Phi__def}
  \Phi(z_1,z_2;{\bf q})& = \frac 1{N^2} \sum_{{\bf k}{\bf k}'}
  G^{(2)}_{{\bf k}{\bf k}'}(z_1,z_2;{\bf q})\ ,\\ \label{eq:Phi_e_def}
  \Phi_{\epsilon}(z_1,z_2;{\bf q})& = \frac 1{N^2} \sum_{{\bf k}{\bf k}'}
  \Delta_{\bf q}\epsilon({\bf k}) G^{(2)}_{{\bf k}{\bf k}'}(z_1,z_2;{\bf
    q})\ ,\\ \label{eq:Phi_ebar_def} \bar{\Phi}_{\epsilon}(z_1,z_2;{\bf
    q})& = \frac 1{N^2} \sum_{{\bf k}{\bf k}'} G^{(2)}_{{\bf k}{\bf
      k}'}(z_1,z_2;{\bf q}) \Delta_{\bf q}\epsilon({\bf k}')\ .
\end{align}
\end{subequations}
The contributions from the two-particle irreducible vertex $\Lambda$ in
Eq.~\eqref{eq:difference_BSE} cancel each other provided the two-particle
irreducible vertex is symmetric, i.~e., $\Lambda_{\mathbf{k}\mathbf{k}'}=
\Lambda_{\mathbf{k}'\mathbf{k}}$.  If so, we end up with a continuity
equation relating the generalized density-density and density-current
correlation functions
\begin{subequations} \label{eq:continuity}
\begin{multline}
  \label{eq:continuity_basic}
  \Phi_{\epsilon}(z_+,z_-;{\bf q}) - \Delta z \Phi(z_+,z_-;{\bf q}) \\ =
  \frac 1N \sum_{\bf k}\Delta_{\bf q} G (\mathbf{k};z_+,z_-) \ .
\end{multline}
Another continuity equation can be derived by multiplying
Eq.~\eqref{eq:difference_BSE} with the energy difference and summing over
the fermionic momenta.  We obtain an equation relating the generalized
current-current and density-current correlation functions
\begin{multline}
  \label{eq:continuity_inter}
  \Phi_{\epsilon\epsilon}(z_+,z_-;{\bf q}) - \Delta z
  \bar{\Phi}_\epsilon(z_+,z_-;{\bf q}) \\ = \frac 1N \sum_{\bf
    k}\Delta_{\bf q} G (\mathbf{k};z_+,z_-)\Delta_{\bf
    q}\epsilon(\mathbf{k})\ .
\end{multline}
Combining the above two continuity equations we obtain the resulting
relation between the generalized current-current and density-density
correlation functions replacing the operator continuity equation
\eqref{eq:continuity_Fourier}
\begin{multline}
\label{eq:continuity_final}
\Phi_{\epsilon\epsilon}(z_+,z_-;{\bf q}) - (\Delta z)^2 \Phi(z_+,z_-;{\bf
  q}) \\ = \frac 1N \sum_{\bf k} \Delta_{\bf q} G
(\mathbf{k};z_+,z_-)\left[\Delta z + \Delta_{\bf q}\epsilon(\mathbf{k})
\right] \ .
\end{multline}
\end{subequations}

Actually, correlation function $\Phi_{\epsilon\epsilon}$ is strictly
speaking not the current-current correlation function, since it is not the
momentum or velocity that is summed with the two-particle Green function.
We used an energy difference $\Delta_{\bf q}\epsilon({\bf k})$ from
Eq.~\eqref{eq:energy_variation}.  Only in the case of quadratic dispersion
relation we have $\Delta_{\bf q}\epsilon({\bf
  k})=\mathbf{q}\cdot\mathbf{k}/m$ and the energy difference is
proportional to the group velocity. Otherwise $\Phi_{\epsilon\epsilon}$
equals the current-current correlation function only in the small-momentum
limit, $q\to0$. We use
\begin{equation}
  \label{eq:small_momentum}
  \Delta_{\bf q}\epsilon({\bf k})  \approx \mathbf{q}\cdot
  \mathbf{v}({\bf k})
\end{equation}
to convert the energy difference in the hydrodynamic limit to a multiple of
the group velocity $\mathbf{v}(\mathbf{k})$ and denote
\begin{multline}
  \label{eq:Phi_vv}
  \Phi_{\alpha\beta}(z_+,z_-;{\bf q})\\ = \frac 1{N^2}\sum_{{\bf k}{\bf
      k}'} v^{\alpha}({\bf k}) v^{\beta}({\bf k}') G^{(2)}_{{\bf k}{\bf
      k}'}(z_+,z_-;{\bf q})\ .
\end{multline}
With the aid of Eq.~\eqref{eq:Phi_vv} we rewrite continuity equation
\eqref{eq:continuity_final} to
\begin{multline}
  \label{eq:continuity_reduced}
  \sum_{\alpha\beta}q^{\alpha}q^{\beta}\Phi_{\alpha\beta} (z_+,z_-;{\bf q})
  - \sum_\alpha q^\alpha\left\langle v^{\alpha} \Delta_{\bf q} G
    (z_+,z_-)\right\rangle\\ = \Delta z \left[\Delta z \Phi(z_+,z_-;{\bf
      q}) + \left\langle \Delta_{\bf q} G (z_+,z_-)\right\rangle \right]
\end{multline}
where the angular brackets stand for the summation over the fermionic
momenta from the first Brillouin zone as defined in
Eq.~\eqref{eq:1P_summed}.  In the isotropic case we obtain
\begin{multline}
  \label{eq:continuity_isotropic}
  \Phi_{\alpha\alpha}(z_+,z_-;{\bf q}) = \frac 1{q^2} \mathbf{q}\cdot
  \left\langle
    \mathbf{v} \Delta_{\bf q} G (z_+,z_-)\right\rangle\\
  + \frac{\Delta z}{q^2}\left[\Delta z \Phi(z_+,z_-;{\bf q}) +
    \langle\Delta_{\bf q}G(z_+,z_-)\rangle\right]\ .
\end{multline}
It is the most general relation between the current-current and
density-density correlation functions for arbitrary complex frequencies and
non-zero momenta. It holds for finite momenta as far as the dispersion law
remains quadratic. For general dispersion relations the validity of
Eq.~\eqref{eq:continuity_isotropic} remains in the asymptotic regime of
small momenta.

To return to measurable quantities we limit the complex frequencies to the
real axis. To derive exact relations we first define a generalization of
the conductivity with the correlation functions
$\Phi_{\epsilon\epsilon}^{ab}$:
\begin{widetext}\begin{multline}
  \label{eq:conductivity_generalized}
  \sigma_{\epsilon\epsilon}(\mathbf{q},\omega) = -e^2
  \int_{-\infty}^{\infty} \frac {dE}{2\pi\omega} \left\{ \left[f(E+\omega)
      - f(E)\right] \Phi_{\epsilon\epsilon}^{AR}(E,E+\omega;\mathbf{q}) +
    f(E)\Phi_{\epsilon\epsilon}^{RR}(E,E+\omega;\mathbf{q}) \right. \\
  \left.  - f(E+\omega) \Phi_{\epsilon\epsilon}^{AA}(E,E+\omega;\mathbf{q})
    - f(E)\left[\Phi_{\epsilon\epsilon}^{RR}(E,E;\mathbf{q}) -
      \Phi_{\epsilon\epsilon}^{AA}(E,E;\mathbf{q})\right]\right\}\ .
\end{multline}
We use continuity equation \eqref{eq:continuity_final} to relate
$\sigma_{\epsilon\epsilon}$ and $\chi$ and find
\begin{multline}
  \label{eq:gauge_invariance}
  \sigma_{\epsilon\epsilon}(\mathbf{q},\omega) + ie^2\omega
  \chi(\mathbf{q},\omega) = \frac {e^2}{\omega}\int_{-\infty}^{\infty}
  \frac {dE}{2\pi} f(E)\left[\Phi^{RR}_{\epsilon\epsilon}(E,E;\mathbf{q}) -
    \Phi^{AA}_{\epsilon\epsilon}(E,E;\mathbf{q})\right] \\ -\frac
  {ie^2}{\omega} \int_{-\infty}^{\infty} \frac {dE}{\pi}f(E) \frac 1N
  \sum_{\mathbf{k}}\Delta_\mathbf{q} \epsilon(\mathbf{k}) \left[\Im
    G^R(\mathbf{k}_+,E) -\Im G^R(\mathbf{k}_-,E)\right] = 0\ .
\end{multline}\end{widetext}
Vanishing of the right-hand side in Eq.~\eqref{eq:gauge_invariance} can be
explicitly manifested when continuity equation \eqref{eq:continuity_final}
is applied to the two-particle correlation functions
$\Phi^{RR}_{\epsilon\epsilon}$ and $\Phi^{AA}_{\epsilon\epsilon}$. The two
terms on the right-hand side of Eq.~\eqref{eq:gauge_invariance} add to
zero.

We can extract the electrical conductivity from
$\sigma_{\epsilon\epsilon}(\mathbf{q},\omega)$ in the limit $q\to0$ or for
quadratic dispersion law by replacing $\Phi_{\epsilon\epsilon}=\sum_\alpha
q_\alpha^2 \Phi_{\alpha\alpha}$ leading to
$\sigma_{\epsilon\epsilon}(\mathbf{q},\omega)=\sum_\alpha q_\alpha^2
\sigma_{\alpha\alpha}(\mathbf{q},\omega)$. In the isotropic case we then
reveal Eq.~\eqref{eq:Einstein_general} from
Eq.~\eqref{eq:gauge_invariance}. It is important to state that this proof
of validity of Eq.~\eqref{eq:Einstein_general} is strongly based on the
Ward identity \eqref{eq:VWW_identity} for nonzero momenta. The latter can
be proved only perturbatively, see Appendix, and hence it is not clear
whether Eq.\eqref{eq:Einstein_general} holds beyond the perturbative regime
near and beyond the Anderson localization transition.

\section{Diffusion, diffusion pole, and Einstein relation in quantum systems}
\label{sec:Diff_Einstein}

In the preceding sections we proved validity of
Eq.~\eqref{eq:Einstein_general} for arbitrary frequencies and small momenta
from the region where quadratic dispersion relation is accurate in strongly
disordered lattice systems. It is a useful relation in particular in
approximate theories. It enables one to approximate only either the density
response function or the conductivity and to determine the other one from
Eq.~\eqref{eq:Einstein_general}. However, both quantities in
Eq.~\eqref{eq:Einstein_general} are response functions that are not well
suited for systematic approximations. We would prefer exact relations or
identities for equilibrium correlation or Green functions. This can be
achieved when we introduce diffusion. Diffusion in classical systems is
defined by Fick's law relating the current with the negative gradient of
the charge density.\cite{Kubo57,Enz92} The diffusion constant introduced in
this way, i.~e. via Fick's law, has in classical physics an intuitive
phenomenological character without a proper microscopic backing that should
come from quantum physics. The study
of diffusion in quantum disordered systems was launched by the seminal work
of Anderson analyzing destructive effects of quantum coherence on diffusion
of disordered electrons.\cite{Anderson58} Since then diffusion has become
part of transport studies of disordered and interacting quantum itinerant
systems.\cite{Altshuler85,Kramer93,Belitz94} However even in these quantum
approaches diffusion is introduced either via (self-consistent)
perturbation expansions leading to the diffusion pole in the electron-hole
correlation function or via the semiclassical limit where the electron-hole
correlation function becomes a Green function of the classical diffusion
equation.\cite{Rammer98}

Here we propose a nonperturbative way to define diffusion from first
principles. We introduce a quantum generalization of Fick's law with a
diffusion response function relating the averaged current density with the
negative gradient of the charge density of a perturbed system. We hence use
\begin{subequations}\label{eq:diffusion}
\begin{multline}
  \label{eq:diffusion_def}
  \left\langle \mathbf{j}(\mathbf{x},t)\right\rangle_{av} = -\ e
  \int_{-\infty}^{\infty} dt' \int d^dx' D(\mathbf{x}-\mathbf{x}',t-t') \\
  \times \boldsymbol{\nabla} \left\langle \delta
    \widetilde{n}(\mathbf{x}',t')\right\rangle_{av}
\end{multline}
as a definition of the diffusion response function $D(\mathbf{x},t)$ being
an extension of the diffusion constant to which it should reduce in the
semiclassical limit. In the above definition we used a perturbed charge
density $\delta \widetilde{n}(\mathbf{x},t)$ that is not the total change
of the charge density $\delta n(\mathbf{x},t)=n(\mathbf{x},t) - n_0$.  It
is a ``dynamical'' density change obeying a condition
\begin{equation}\label{eq:density_conservation}
\int_{-\infty}^{\infty} dt \langle\delta
\widetilde{n}(\mathbf{x},t)\rangle_{av}=0
\end{equation}
\end{subequations}
saying that the total change at any place during the entire time evolution
is zero. This restriction is dictated by the same condition obeyed by the
averaged current density defined from the conductivity
\eqref{eq:conductivity}.

Definition of diffusion \eqref{eq:diffusion} is nonperturbative, has
quantum character and does not demand any assumption about the behavior in
the semiclassical limit or the existence of the diffusion pole in the
electron-hole correlation function. Moreover, it offers a microscopic
determination of the diffusion function from first principles, i.~e.,
quantum response functions of the perturbed system and via the Kubo
formalism from correlation and Green functions of an equilibrium solution.
It is hence directly related to measurable quantities.

In Fourier representation we obtain for the averaged current
\begin{equation}
  \label{eq:diffusion_Fourier}
  \left\langle \mathbf{j}(\mathbf{q},\omega\right\rangle_{av} = - i e
  \mathbf{q} D(\mathbf{q},\omega)  \left\langle \delta
    \widetilde{n}(\mathbf{q},\omega)\right\rangle_{av} \ .
\end{equation}
The dynamical charge density $\delta \widetilde{n}$ is induced by an
external dynamical scalar potential, which can be expressed in Fourier
representation as
\begin{equation}
  \label{eq:density_prime}
   \left\langle \delta \widetilde{n}(\mathbf{q},\omega)\right\rangle_{av} =
   -\ e\left[\chi(\mathbf{q},\omega) - \chi(\mathbf{q},0)\right]
\varphi(\mathbf{q},\omega)\ .
\end{equation}
It is easy to verify that with this definition
Eq.~\eqref{eq:density_conservation} is fulfilled.  We use
Eq.~\eqref{eq:induced_current} together with
Eqs.~\eqref{eq:diffusion_Fourier} and \eqref{eq:density_prime} to find an
identity between the conductivity, diffusion and density response
\begin{equation}
  \label{eq:C_D_DR}
  \sigma(\mathbf{q},\omega) = -e^2 D(\mathbf{q},\omega)
  \left[\chi(\mathbf{q},\omega) - \chi(\mathbf{q},0)\right] \ .
\end{equation}
We have derived a general relation between quantum diffusion and the
electrical conductivity. It is now to show that quantum diffusion response
function $D(\mathbf{q},\omega)$ behaves in the semiclassical limit as
expected from the diffusion constant in the diffusion equation. Namely, we
have to show that the homogeneous static diffusion $D(\mathbf{0},0)$ obeys
Einstein's relation and enters the diffusion equation in the semiclassical
limit. To come to it we use Eq.~\eqref{eq:Einstein_general} to exclude the
density response $\chi(\mathbf{q},\omega)$ from Eq.~\eqref{eq:C_D_DR} and
Eq.~\eqref{eq:DRF_static} to determine $\chi(\mathbf{q},0)$. Doing so we
obtain in the isotropic case
\begin{multline}
  \label{eq:diff_cond}
  \left[1 + \frac{iq^2}{\omega}D(\mathbf{q},\omega)\right]
  \sigma(\mathbf{q},\omega)\\ = e^2 D(\mathbf{q},\omega)
  \int_{-\infty}^{\infty} \frac{dE}{\pi} f(E) \Im\Phi^{RR}_E(\mathbf{q},0)\
  .
\end{multline}
If we define the homogeneous dynamical conductivity $\sigma(\omega)=
\sigma(\mathbf{0},\omega)$ (in the same way also the dynamical diffusion)
and use the Velick\'y-Ward identity \eqref{eq:VW_momentum} in the limiting
case $z_1-z_2\to0$ we obtain a dynamical generalization of the well known
Einstein relation
\begin{subequations}\label{eq:Einstein_new}
  \begin{multline}
    \label{eq:Einstein_T}
    \sigma(\omega) = e^2 D(\omega) \int_{-\infty}^{\infty} \frac{dE}{\pi}
    f'(E)\Im G^R(E)\\ = e^2 D(\omega) \left(\frac{\partial n}{\partial
        \mu}\right)_T
  \end{multline}
  that at zero temperature reduces to
\begin{equation}
  \label{eq:Einstein_0}
  \sigma(\omega) = e^2 n_F D(\omega) \ .
\end{equation}
\end{subequations}
A classical version of this formula was proved by Einstein for the Brownian
motion of a particle in a random medium\cite{Einstein56} and later on it
was re-derived in the framework of nonequilibrium statistical mechanics by
Kubo.\cite{Kubo57} Here we derived Einstein's relation for quantum response
functions and showed that it is a consequence of gauge invariance of the
system, namely Eq.~\eqref{eq:Einstein_general}.

It is not yet clear from the above reasoning how the diffusion function
$D(\mathbf{q},\omega)$ is related to the classical diffusion equation and
to the diffusion pole.  The existence of the diffusion pole in quantum
systems is usually deduced from the semiclassical limit leading to the
diffusion equation.  We can prove the existence of the diffusion pole
entirely from first quantum principles and from the quantum generalization
of Fick's law, Eq.~\eqref{eq:diffusion}.  We use
Eq.~\eqref{eq:Einstein_general} to exclude now conductivity
$\sigma(\mathbf{q},\omega)$ from Eq.~\eqref{eq:C_D_DR}. It is easy to come
to a representation of the density response function
\begin{equation}
  \label{eq:density_diffusion}
  \chi(\mathbf{q},\omega) = \frac {D(\mathbf{q},\omega)\chi(\mathbf{q},0)
    q^2}{-i\omega + D(\mathbf{q},\omega) q^2}
\end{equation}
known from other treatments of diffusion.\cite{Forster90}
Eq.~\eqref{eq:density_diffusion} holds for arbitrary frequencies and
momenta within the range of quadratic dispersion relation. This
representation directly indicates a singularity in the density response
function in the limit $\omega\to0,q\to0$. Since the order of the limits is
relevant for the result, we have to specify it explicitly.  To single out
the singular contribution in the density response function we have to
choose $\omega/q\ll1$, that is, the inverse situation to
Sec.~\ref{sec:Einstein_relation} and to Eqs.~\eqref{eq:Einstein_new}.

The singularity in the density response function leads to a pole in the
two-particle Green function. It is demonstrated in the low-energy and
small-momentum asymptotics of the density response function at zero
temperature.  We find from Eq.~\eqref{eq:DRF_real}
\begin{multline}
  \label{eq:DRF_0}
  \chi(\mathbf{q},\omega) = \chi(\mathbf{q},0) + \frac{i\omega}{2\pi}
  \left( \Phi^{AR}_{E_F}(\mathbf{q},0) + O(q^0)\right)\\ + O(\omega^2) \ .
\end{multline}
Using this asymptotics in Eq.~\eqref{eq:density_diffusion} we obtain an
explicit manifestation of the diffusion pole in the zero-temperature
electron-hole correlation function
\begin{equation}
  \label{eq:Phi_diffusion}
  \Phi^{AR}_{E_F}({\bf q},\omega)\approx  \frac {2\pi n_F}{-i\omega +
    Dq^2} \
\end{equation}
representing the singular part of the electron-hole correlation function.
We denoted a diffusion constant $D=\lim_{q\to0}D(\mathbf{q},0)$.  This
representation was derived for zero temperature in the asymptotic limit
$q\to0,\omega\to0$ with the restriction $\omega/q\ll1$. To extend it to the
opposite limit, $q/\omega\ll1$, we have to show that the diffusion function
$D(\mathbf{q},\omega)$ is analytic in the limit $q\to0,\omega\to0$.  It is
not a priori clear that the diffusion constant in Eq.~\eqref{eq:Phi_diffusion}
equals the diffusion constant obtained from the inverse order of limits
$D=\lim_{\omega\to0}\lim_{q\to0} D(\mathbf{q},\omega)$ used in the Einstein
relation, Eq.~\eqref{eq:Einstein_0}. However, the diffusion response
function $D(\mathbf{q},\omega)$ was introduced in such a way that the
limits $\omega\to0$ and $q\to0$ commute and we have only one definition for
the static diffusion constant from the density response function
\begin{multline}\label{eq:diff_const}
  D = \frac i2\ \frac{\lim_{\omega\to0}\nabla^2_q
    \chi(\mathbf{q},\omega)|_{q=0}}
  {\lim_{q\to0}\partial_\omega\chi(\mathbf{q},\omega)|_{\omega=0}}\\ =
  \frac {2\pi n_F}{\lim_{q\to0}q^2\Phi^{AR}(\mathbf{q},0)}\ .
\end{multline}
The left equality in Eq.~\eqref{eq:diff_const} holds quite generally while
the right one only at zero temperature. Note that representation
\eqref{eq:Phi_diffusion} holds only at zero temperature for small momenta
and in the low-frequency limit.

We can see from Eq.~\eqref{eq:Phi_diffusion} that the diffusion function
introduced in the quantum version of Fick's law reduces to the diffusion
constant determining the long-range fluctuations of the electron-hole
correlation function and hence in the semiclassical limit also the
diffusive behavior in the diffusion equation. However, only the static
diffusion constant $D$ enters the denominator of the electron-hole
correlation function.  A generalization of the diffusion constant from the
electron-hole correlation function to a frequency-dependent quantity via
\begin{equation}
  \label{eq:AR_diffusion}
  n_F\widetilde{D}(\omega) = \frac{\omega^2}{4\pi} \nabla^2_{q}
  \Phi^{AR}_{E_F}(\mathbf{q},\omega)\big|_{q=0}  \ ,
\end{equation}
does not have a rigorous justification.  It is nevertheless often used in
the literature for a quantitative treatment of Anderson localization.\cite{Vollhardt92}
Beware that the dynamical diffusion defined in this way does not fulfill
the Einstein relation and hence deviates from the dynamical conductivity.
We show in the next section that in an exactly solvable mean-field limit
the left-hand side of Eq.~\eqref{eq:AR_diffusion} related to the
conductivity via the Einstein relation differs at finite frequencies from
the conductivity obtained from the Kubo formula with the electrical
current.

\section{Infinite-dimensional model: Explicit exact solution}
\label{sec:infinite_d}

We derived exact relations between density and current correlation
functions for a lattice electron gas in a random potential. We now
demonstrate the generally derived formulas explicitly on an exactly
solvable limit of infinite spatial dimensions. This limit serves as a
mathematical tool for the definition of a mean-field theory not only for
classical spin systems but also for itinerant disordered and interacting
models.\cite{Janis89,Janis92,Georges96} In case of the Anderson model of
disordered electrons the mean-field theory, i.~e., the limit of infinite
spatial dimensions, equals the coherent potential approximation. We hence
resort in this section to this solution.

Equation determining the local self-energy in the coherent-potential
approximation is Soven's equation that can be written as\cite{Janis01b}
\begin{subequations}\label{eq:CPA_equations}
 \begin{equation}\label{eq:CPA_1}
 G(z)=\left\langle\left[G^{-1}(z) + \Sigma(z) -
V_i\right]^{-1}\right\rangle_{av}
 \end{equation}
 where $G(z)=N^{-1}\sum_\mathbf{k}G(\mathbf{k},z)$. The two-particle
 irreducible vertex then is
\begin{widetext} \begin{multline}\label{eq:CPA_2}
    \lambda(z_+,z_-)=\frac 1{G(z_+)G(z_-)}\left[1- {\left\langle \frac 1{
            1+\left(\Sigma(z_+)-V_i\right)G(z_+)}\frac 1{ 1+\left(
              \Sigma(z_-)-V_i\right)G(z_-)}\right\rangle_{av}}^{-1}
    \right]\\ = \frac {\Sigma(z_+) - \Sigma(z_-)} {G(z_+) - G(z_-)} = \frac
    {\Delta\Sigma} {\Delta G} \ .
 \end{multline}\end{widetext}
\end{subequations}
The angular brackets were defined in Eq.~\eqref{eq:1P_summed}. The second
equality in Eq.~\eqref{eq:CPA_2} is the Velick\'y-Ward identity.

It is straightforward to find an explicit form of the two-particle Green
function in the coherent-potential approximation
\begin{multline}
  \label{eq:CPA_2PGF}
  G^{(2)}_{\mathbf{k}\mathbf{k}'}(z_+,z_-;\mathbf{q}) = G_+(\mathbf{k})
  G_-(\mathbf{k})\left[\delta(\mathbf{k} - \mathbf{k}')\right. \\ \left. +\
    \frac {\lambda(z_+,z_-)G_+(\mathbf{k}') G_-(\mathbf{k}')} {1 -
      \lambda(z_+,z_-)\langle G_+G_-\rangle}\right]
\end{multline}
where $G_\pm(\mathbf{k})=G(\mathbf{k}\pm\mathbf{q}/2,z_\pm)$ and $\langle
G_+G_-\rangle=N^{-1}\sum_{\mathbf{k}}G_+(\mathbf{k}) G_-(\mathbf{k})$.

Using the two-particle Green function, Eq.~\eqref{eq:CPA_2PGF}, we easily
obtain explicit representations for the density-density and density-current
correlation functions
\begin{subequations}\label{eq:CPA_Phi}
\begin{align}
    \label{eq:CPA_Phi_0}
    \Phi(z_+,z_-;\mathbf{q}) = \frac {\langle G_+G_-\rangle} {1
      - \lambda(z_+,z_-)\langle G_+G_-\rangle}\ , \\
\label{eq:CPA_Phi_e}
\Phi_\epsilon(z_+,z_-;\mathbf{q}) = \frac {\langle
  \Delta_\mathbf{q}\epsilon G_+G_-\rangle} {1 - \lambda(z_+,z_-)\langle
  G_+G_-\rangle}\ .
\end{align}
\end{subequations}

We now insert Eqs.~\eqref{eq:CPA_Phi} together with the explicit
representation for the two-particle Green function,
Eq.~\eqref{eq:CPA_2PGF}, to the right-hand side of
Eq.~\eqref{eq:continuity_basic} and find
\begin{multline}
  \label{eq:CPA_continuity}
  \Phi_\epsilon(z_+,z_-;\mathbf{q}) -\Delta z \Phi(z_+,z_-;\mathbf{q})\\ =
  \frac {\left\langle\left(\Delta_{\mathbf{q}}\epsilon -\Delta
        z\right)\frac{\Delta_{\mathbf{q}}G}{\Delta_{\mathbf{q}}\epsilon -
        \Delta z -\Delta\Sigma}\right\rangle} {1 -
    \lambda(z_+,z_-)\langle G_+ G_-\rangle}\\
  = \frac {\langle \Delta G\rangle - \Delta\Sigma\langle G_+ G_-\rangle}{1
    - \lambda(z_+,z_-)\langle G_+ G_-\rangle} = \langle \Delta G\rangle\ ,
\end{multline}
that is, continuity equation \eqref{eq:continuity_basic} is fulfilled.
Analogously we can explicitly verify Eqs.~\eqref{eq:continuity_inter} and
\eqref{eq:continuity_final}. Actually, validity of continuity equations
\eqref{eq:continuity} follows from the Velick\'y-Ward identity, since the
two-particle irreducible vertex $\lambda(z_+,z_-)$ is momentum independent.
In such a case the Vollhardt-W\"olfle and the Velick\'y identities are
equivalent.

As next we show that the $dc$-conductivity in the coherent-potential
approximation can be calculated using the diffusion constant $D$ from
Eq.~\eqref{eq:Phi_diffusion}. To this purpose we need to evaluate
$\nabla^2_{q}\langle G_+G_-\rangle$ at zero transfer frequency $\omega=0$.
We have
\begin{multline}
  \label{eq:nabla_bubble}
  \nabla^2_{q}\langle G_+G_-\rangle = \frac 12 \nabla_q\langle\left(G_-
    \nabla_k G_+ - G_+\nabla_k G_-\right)\rangle\\ = \frac 14 \langle G_-
  \nabla^2_k G_+ + G_+\nabla^2_k G_- - 2\nabla_k G_+ \nabla_k G_- \rangle\\
  = - \langle \nabla_k G_+ \nabla_k G_-\rangle
\end{multline}
where in the last equality we used integration per parts in momentum space.
We utilize this result and Eq.~\eqref{eq:VW_momentum} to obtain the
electron-hole correlation function
\begin{equation}
\label{eq:CPA_eh}
\Phi(\mathbf{q},\Delta z) = - \frac {\Delta G}{\Delta z - \displaystyle
  \frac 12  \frac {\Delta\Sigma}{\Delta\Sigma- \Delta z}\frac {\langle
    v_k^2(\Delta G)^2 \rangle}{\Delta G}q^2}\ ,
\end{equation}
that in the low-energy limit reduces for real frequencies to
\begin{equation}
\label{eq:CPA_diff}
 \Phi^{AR}(\mathbf{q},\omega) = \frac {2\pi n_F}{-i\omega
   +\sigma^{CPA}q^2/e^2n_F}
\end{equation}
proving the Einstein relation for the $dc$-conductivity $\sigma^{CPA}$ of
the coherent-potential approximation. It is evident from
Eq.~\eqref{eq:CPA_eh} that the frequency-dependent coefficient in the
denominator of the electron-hole correlation function does not equal the
dynamical diffusion from Eq.~\eqref{eq:Einstein_0}.  To show this
explicitly we use definition \eqref{eq:AR_diffusion} and compare the result
with the real-part of the conductivity at zero temperature given by
\begin{subequations}\label{eq:CPA_conductivity}
\begin{multline}\label{eq:CPA_cond_0}
  \Re\sigma_{\alpha\alpha}(\omega) = e^2\int_{E_F-\omega}^{E_F} \frac{d E}
  {2\pi\omega} \left[\Re\Phi^{AR}_{\alpha\alpha}(E,E+\omega;\mathbf{0})
  \right. \\ \left. -\
    \Re\Phi^{RR}_{\alpha\alpha}(E,E+\omega;\mathbf{0})\right] =
  \sigma^{AR}_{\alpha\alpha}(\omega) + \sigma^{RR}_{\alpha\alpha}(\omega)\
  .
\end{multline}
We separated contributions to the conductivity from the electron-hole and
electron-electron current-current correlation functions, $\sigma^{AR}$,
$\sigma^{RR}$. It is useful to introduce also a simplified asymptotic form
of the low-frequency conductivity
\begin{multline}\label{eq:CPA_cond_lin}
  \Re\bar{\sigma}_{\alpha\alpha}(\omega) = e^2\frac{1} {2\pi}
  \left[\Re\Phi^{AR}_{\alpha\alpha}(E_F-\omega,E_F;\mathbf{0}) \right. \\
  \left. -\ \Re\Phi^{RR}_{\alpha\alpha}(E_F-\omega,E_F;\mathbf{0})\right]
\end{multline}
\end{subequations}
where the integrand in the energy integral is replaced by its initial
value. This representation is asymptotically exact in the limit
$\omega\to0$ for smooth current-current correlation functions
$\Phi_{\alpha\alpha}$ near the Fermi energy $E_F$ and carries the same
frequency dependence as the diffusion constant $\widetilde{D}(\omega)$
defined in Eq.~\eqref{eq:AR_diffusion}.

\begin{figure}
  \resizebox{7.5cm}{!}{\includegraphics{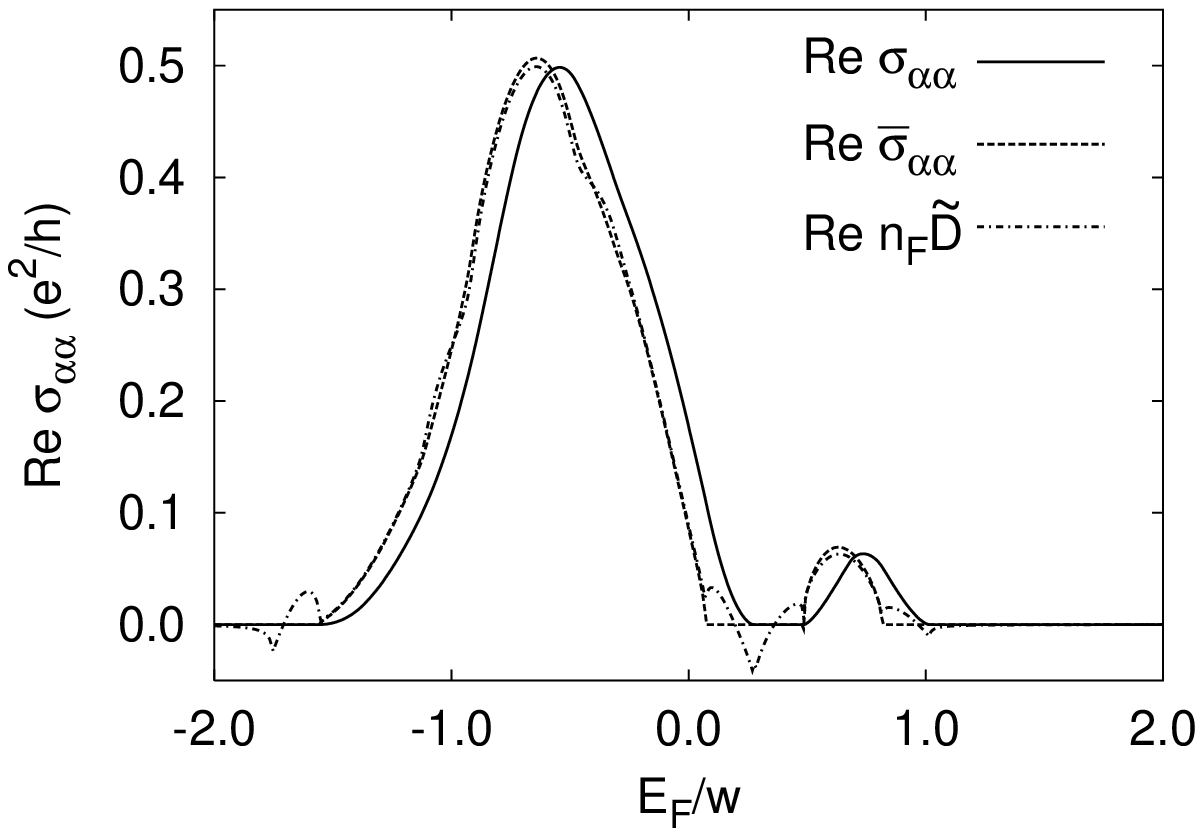}}
  \resizebox{7.5cm}{!}{\includegraphics{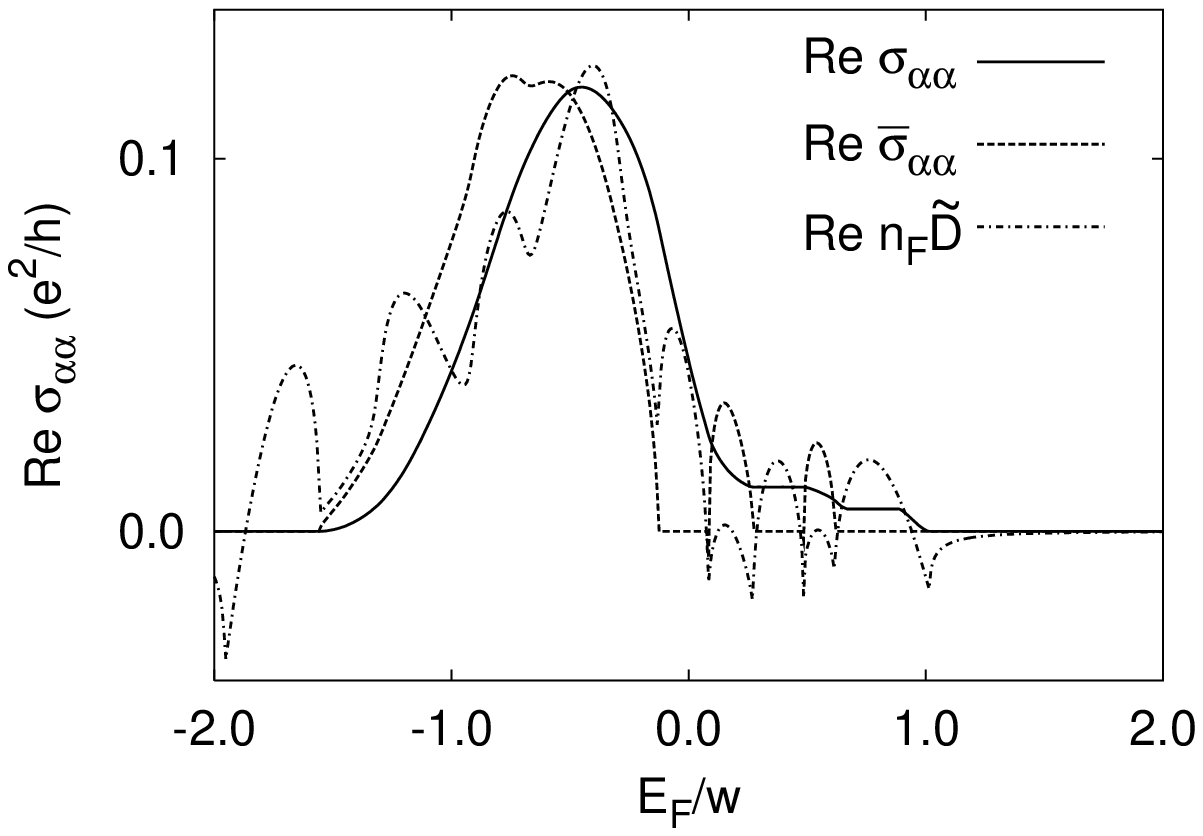}}
\caption{\label{fig:om_01} Dynamical conductivities defined in three
  different manners, Eqs.~\eqref{eq:AR_diffusion},
  \eqref{eq:CPA_conductivity} for a binary alloy on a simple cubic lattice
  with the random potential $V_i=\pm0.6w$, weighted with $x=0.1$ and $1-x$,
  respectively. Frequency is $\omega=0.2w$ in the upper pane and
  $\omega=0.4w$ in the lower pane.  Here $w$ is the energy half-bandwidth.}
\end{figure}

For the numerical calculations we use a binary alloy with two values of the
random potential $V_i=\pm\Delta$ weighted with probability $x$ and $1-x$.
Fig.~\ref{fig:om_01} shows the two dynamical conductivities and the
diffusion constant from Eq.~\eqref{eq:AR_diffusion} on a simple cubic
lattice with parameters $x=0.1$, $\Delta=0.6w$ for frequencies
$\omega=0.2w$ and $\omega=0.4w$ with $w$ being the half-bandwidth. Note
that the split-band value of the disorder strength is
$\Delta_c\approx0.4w$.
Although the three quantities differ for low frequencies only slightly
inside the band, they behave differently near the band edges and for higher
frequencies. In particular the diffusion constant shows anomalous behavior
when the Fermi energy approaches a band edge. The compensating terms from
the electron-electron (density-density) correlation function get relevant
there.  Anomalous behavior of the diffusion constant is more transparent
for higher frequencies (lower pane).  We cannot evidently rely on
$n_F\widetilde{D}(\omega)$ as a good approximation to the conductivity for
finite frequencies except for Fermi energies deep inside the energy band.
The smoothing impact of the integral over frequencies on the behavior of
the dynamical conductivity gets clear from our numerical results.

\begin{figure}
  \resizebox{7.5cm}{!}{\includegraphics{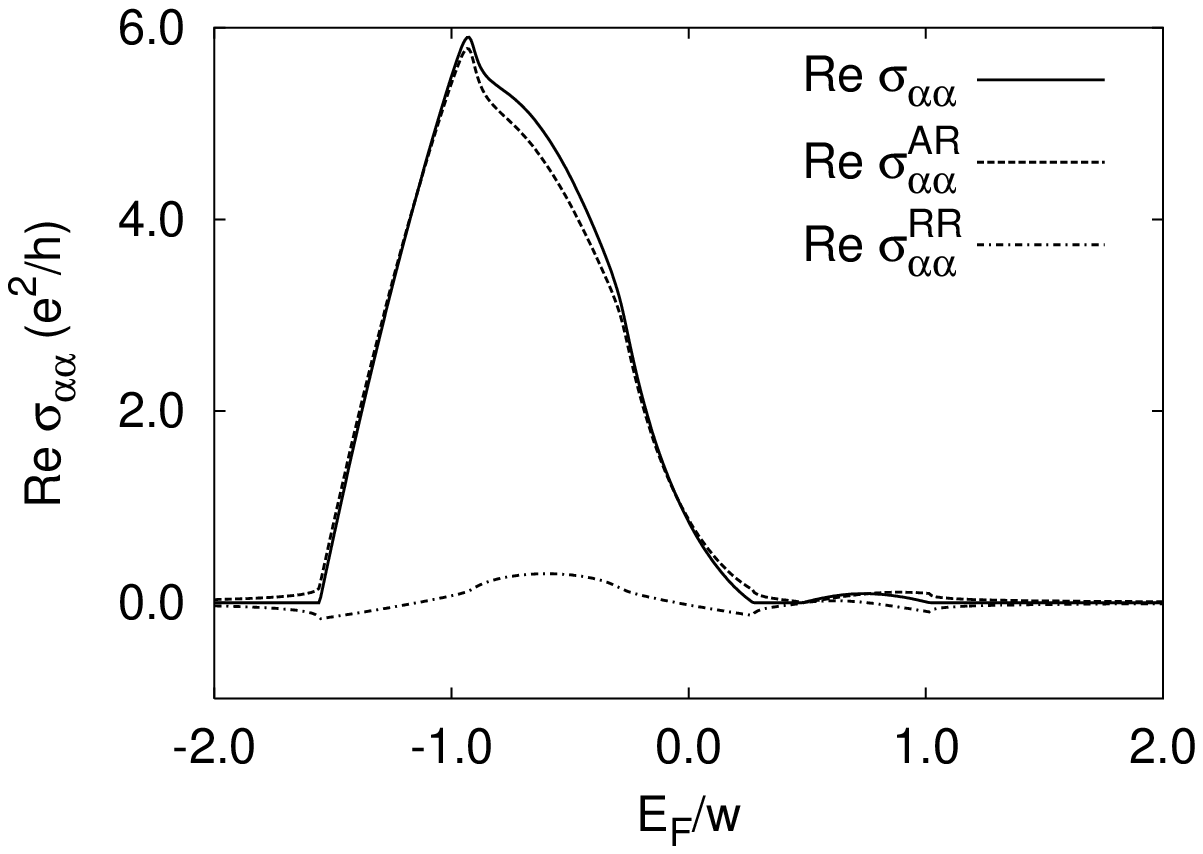}}
  \resizebox{7.5cm}{!}{\includegraphics{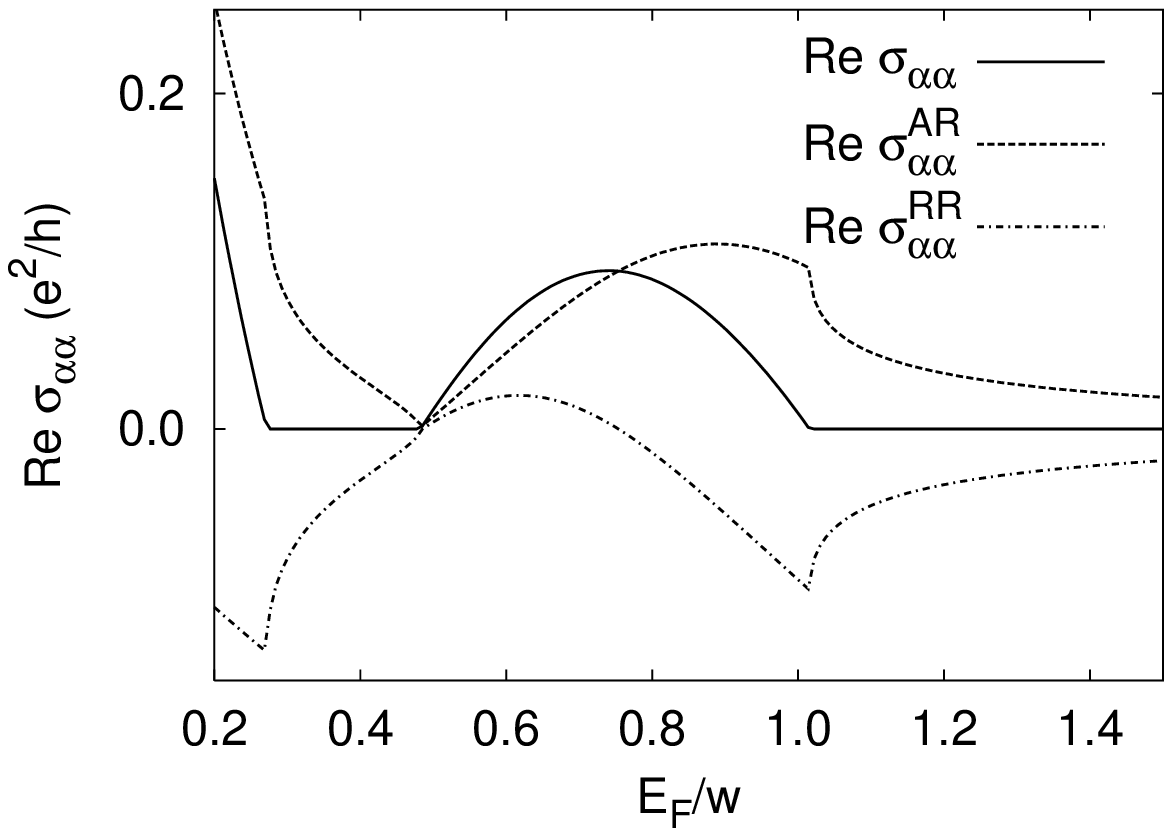}}
\caption{\label{fig:sigma_AR_0} Weight of contributions to the full
  conductivity from the electron-hole and electron-electron parts in the
  static case, $\omega=0$. The setting is the same as in
  Fig.~\ref{fig:om_01}. The lower pane shows the details of the satellite
  (split-off) band.}
\end{figure}

\begin{figure}
  \resizebox{7.5cm}{!}{\includegraphics{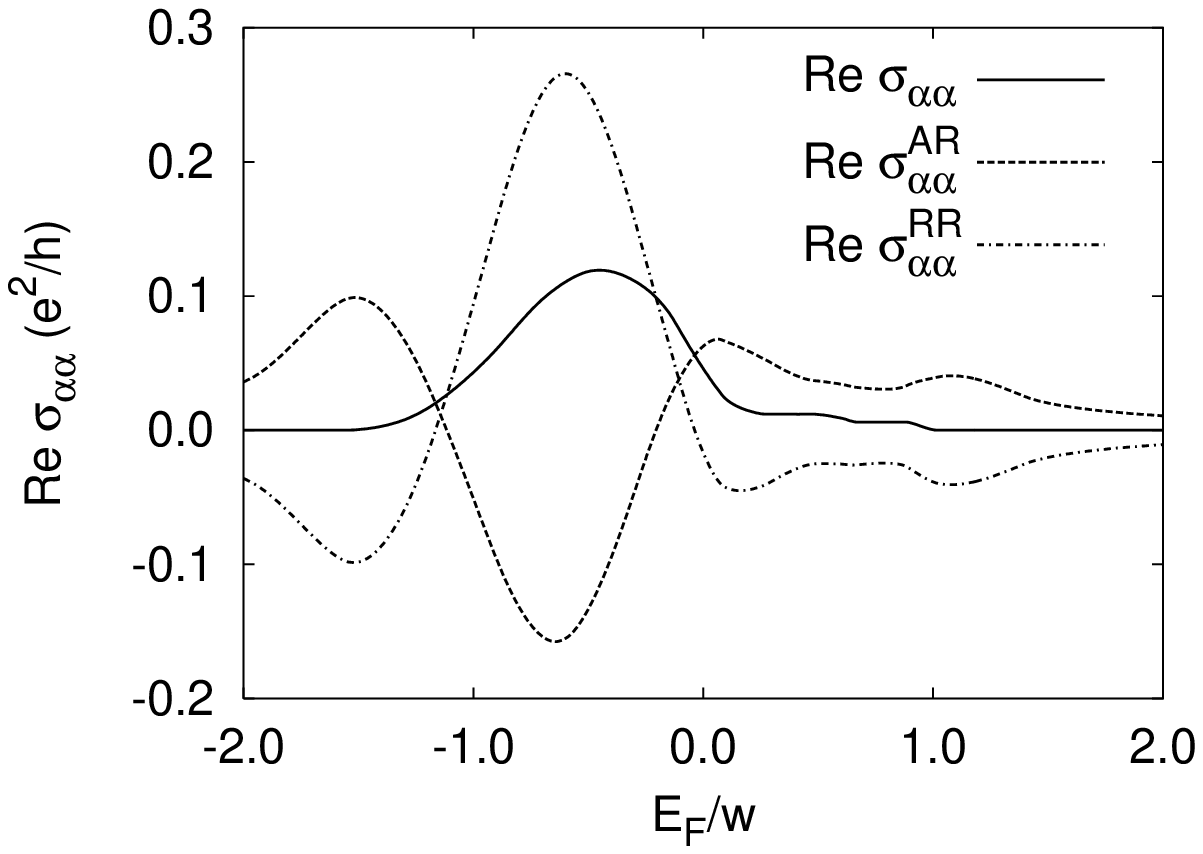}}
  \resizebox{7.5cm}{!}{\includegraphics{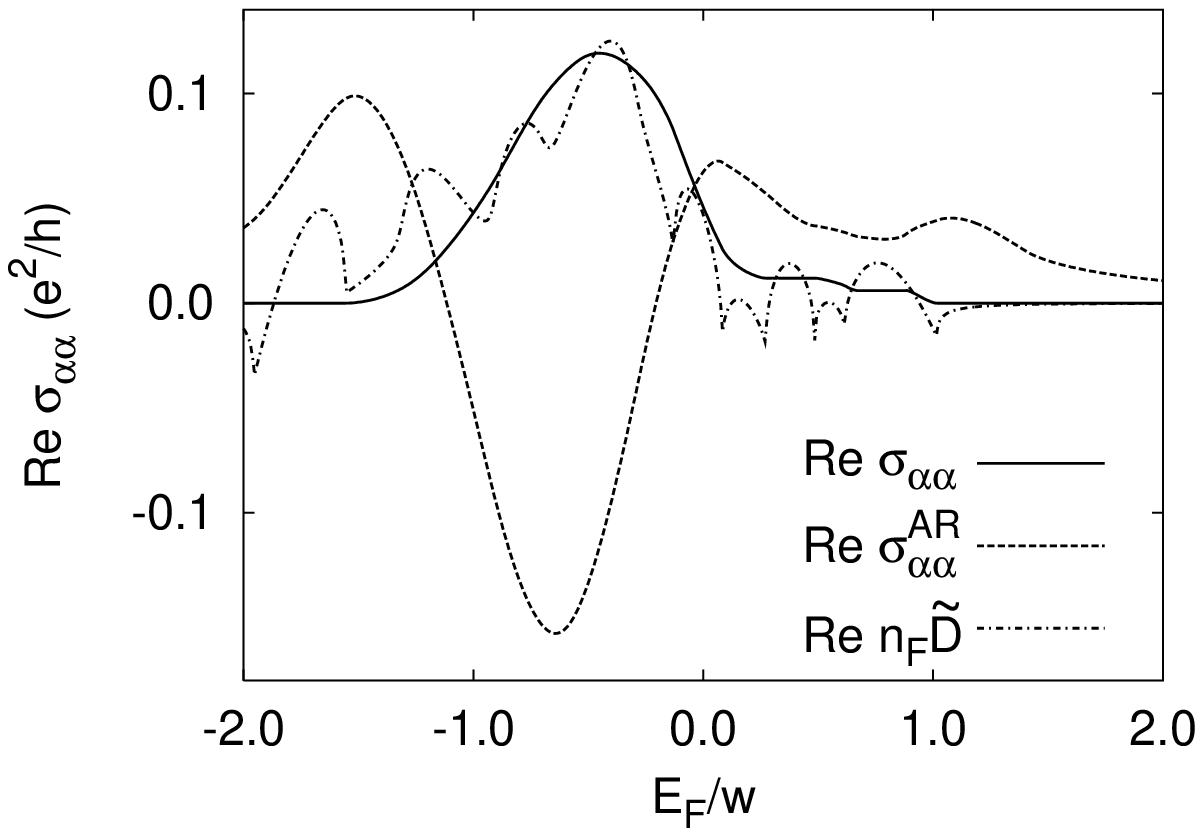}}
\caption{\label{fig:sigma_AR_1} Electron-hole and electron-electron
  contributions to the conductivity at $\omega=0.4w$. The lower pane shows
  differences between the conductivity obtained from the diffusion constant
  $\widetilde{D}$ and $\sigma^{AR}$.}
\end{figure}

Recently a discussion was renewed about the proportion of contributions to
the Kubo formula for the electrical conductivity from the electron-hole and
electron-electron current-current correlation functions, $\sigma^{AR}$ and
$\sigma^{RR}$ in Eq.~\eqref{eq:CPA_cond_0}, respectively.\cite{Nikolic01}
In Fig.~\ref{fig:sigma_AR_0} these contributions are compared for the same
setting of the binary alloy on a simple cubic lattice for $\omega=0$. The
contribution from $\sigma^{AR}$ dominates inside the band far from the band
edges. Outside the central band and in the satellite impurity band the
importance of the compensating effects of $\sigma^{RR}$ is evident.
The situation worsens when we go over to the dynamical conductivity,
Fig.~\ref{fig:sigma_AR_1}. We can see that there is no region where the
electron-hole contribution would dominate or approximate the full
conductivity reliably.
It is important to note that conductivity $\sigma^{AR}$ differs from the
(dynamical) diffusion constant defined from the electron-hole
density-density correlation function, Eq.~\eqref{eq:AR_diffusion}. The
latter contains even in the static limit both terms, $\sigma^{AR}$ and
$\sigma^{RR}$. The difference is evident in Fig.~\ref{fig:sigma_AR_1}.  The
conductivity calculated from $\widetilde{D}(\omega)$ is much closer to the
full electrical conductivity than $\sigma^{AR}$. With increasing frequency
the role of the term $\sigma^{RR}$ increases. We hence cannot interchange
contributions to the electrical conductivity from the electron-hole
density-density correlation function $\Phi^{AR}(\mathbf{q},\omega)$ (its
leading $q$-dependent term) and the conductivity
$\sigma^{AR}(\mathbf{q},\omega)$. The former is generally better
approximation than the latter except for Fermi energies near band edges and
in band tails where both the approximations equally fail.

\section{Discussion and conclusions}
\label{sec:conclusions}

In this paper we derived a number general relations between
density-density, density-current and current-current correlation functions.
With the aid of these relations we proved validity of an equation relating
the electrical conductivity to the density response function,
Eq.~\eqref{eq:Einstein_general}, in the hydrodynamic limit of strongly
disordered electron lattice systems and derived a dynamical generalization
of the Einstein relation between the conductivity and diffusion,
Eq.~\eqref{eq:Einstein_new}. We found to what extent
Eq.~\eqref{eq:Einstein_general} can be used in quantitative calculations
and when it may be broken. We showed that the dynamical diffusion from the
Einstein relation cannot be related to the diffusion pole beyond the static
limit.  We used only exact reasoning and equalities derived directly from
equations of motion for Green functions. We avoided extrapolations of
semiclassical results or conclusions derived within equilibrium theory.

The electrical conductivity can be related to the density response function
only if Ward identities are fulfilled. The same holds for the existence of
a diffusion pole in the electron-hole correlation function with the static
optical conductivity governing its long-range spatial fluctuations.  There
are two, not fully equivalent, Ward identities for noninteracting electrons
in a random potential. Although they reflect conservation laws they can be
proved only if perturbation expansion in the strength of the random
potential for the irreducible two-particle vertex converges and results in
an analytic function.

In Sec.~\ref{sec:Einstein_relation} we used Velick\'y-Ward identity
\eqref{eq:VW_momentum} to relate mobility with the density response.
This Ward identity expresses conservation of probability in the state
space and holds only for the homogeneous case, i.~e., zero transfer
momentum in the two-particle Green function.  To be of use in the
calculation of the conductivity we had to assume analyticity of the
hydrodynamic limit for all frequencies and to use the transfer
momentum as an expansion parameter.  Vollhardt-W\"olfle-Ward identity
\eqref{eq:VWW_identity} used in Sec.~\ref{sec:dynamical_conductivity}
to reveal the density response function from the conductivity
expresses charge conservation.  It holds, unlike the Velick\'y-Ward
identity, for arbitrary transfer momenta in the two-particle function,
but it can be proved only via a perturbation (diagrammatic) expansion
for the two-particle vertex function. Moreover, to warrant this
identity we are not allowed to sum selected classes of relevant
diagrams.  A single relevant diagram for the two-particle irreducible
vertex of order $n$ produces $n-1$ irrelevant diagrams in the proof of
the Ward identity. Selection of relevant diagrams may be, however,
dictated by the existence of a singularity in the two-particle
function or by analyticity (causality) of the theory. There is hence
no general guarantee that Ward identities and consequently the
relation between the conductivity and the density response hold beyond
the perturbative regime near the Anderson localization transition.

When the conductivity can be related to the density response via
Eq.~\eqref{eq:Einstein_general} we can derive a quantum version of the
Einstein relation. To that purpose we introduced a diffusion response
function relating the averaged current with the negative gradient of the
charge density as a quantum generalization of Fick's law. Diffusion defined
in this way enters the dynamical Einstein relation,
Eq.~\eqref{eq:Einstein_new}, but it is not \`a priori clear whether this
quantum diffusion reduces to the classical one in the semiclassical limit.
The solution of the classical diffusion equation pops up from a quantum
theory via the diffusion pole and the long-range asymptotics of the
electron-hole correlation function. We showed that the ``classical''
diffusion from the diffusion pole equals the quantum one from the Einstein
relation and hence the conductivity only in the static limit.  A dynamical
diffusion constant defined from the electron-hole correlation function is
no longer proportional to the conductivity beyond the static limit. This
was demonstrated generally as well as explicitly on an exact solution in
infinite spatial dimensions.

To conclude, we showed that gauge invariance of an electron gas in a random
potential responding to an electromagnetic perturbation is guaranteed
within linear-response theory if Ward identities \eqref{eq:VW_momentum} and
\eqref{eq:VWW_identity} are fulfilled. In this case an (approximate)
averaged two-particle Green function generates density and current response
functions consistent in the hydrodynamic limit with relation
\eqref{eq:Einstein_general}. Or, what is more important, we can safely
apply the approximate two-particle Green function only in either Kubo
formula and use Eq.~\eqref{eq:Einstein_general} to determine the other
response function. We explicitly demonstrated that Ward identities are
fulfilled only in the ``diffusive'' regime characterized by convergence of
the perturbation expansion for the two-particle irreducible vertex and by
analyticity of the hydrodynamic limit for all frequencies. We, however,
know that in solutions with weakly or strongly localized electrons a pole
in the two-particle irreducible vertex emerges and validity of the Ward
identities cannot be generally proved. One has to bear this aspect in mind
when interpolating between the diffusive and localized regimes.

\begin{acknowledgments}
  This work was supported in part by Grant No.  202/01/0764 of the Grant
  Agency of the Czech Republic.
\end{acknowledgments}

\appendix*
\section{Proof and validity of the Vollhardt-W\"olfle-Ward identity}

Vollhardt-W\"olfle-Ward identity \eqref{eq:VWW_identity} is a primary tool
for proving the continuity equations relating the current and density
correlation functions. Understanding its proof and the domain of its
validity is hence very important. The original proof of Vollhardt and
W\"olfle\cite{Vollhardt80b} is restricted to real frequencies with
different small imaginary parts only. However, it is valid for arbitrary
complex frequencies. To demonstrate this we summarize the principal
assumptions and steps of its proof from which we conclude on the domain of
validity of Eq.~\eqref{eq:VWW_identity}.

First, we have to assume that the self-energy $\Sigma$ is a functional of
the full one-electron propagator $G$
\begin{equation}
  \label{eq:Sigma_def}
  \Sigma({\bf p},z) = \Sigma\left[G\right]({\bf p},z)
\end{equation}
so that the perturbation expansion in the random potential consists of
irreducible diagrams only. We consider here only noninteracting electrons
in a random potential although the reasoning can be extended to correlated
electrons as well. Perturbation expansion of the self-energy $\Sigma({\bf
  p},z)$ in the random potential reads
\begin{widetext}\begin{multline}
\label{eq:Sigma_expansion} \Sigma({\bf p},z) = \langle
V\rangle_{av} + \sum_{n=1}^\infty \sum_{k=1}^{[(n+1)/2]}
\sum^{n+1}_{j_1,\ldots, j_{k}=1} \sum_{l_1,\ldots,l_k}\frac 1{l_1!\ldots
  l_k!}  \delta(l_1 + \ldots l_k-n-1) \left\langle
  V_{j_1}^{l_1}\right\rangle_c
\cdots \left\langle V_{j_k}^{l_k}\right\rangle_c \\
\sum_{{\bf R}_1,\ldots,{\bf R}_{n+1}}
\sum_{\mathcal{P}\{i_1,\ldots,i_{n+1}\}} \delta_{j_1,i_1} \ldots
\delta_{j_1,i_{l_1}} \delta_{j_2,i_{l_1+1}}\ldots
\delta_{j_k,i_{n+1}}\\
\frac 1{N^n} \sum_{{\bf p}_1,\ldots,{\bf p}_n} e^{i{\bf R}_{i_1}({\bf p} -
  {\bf p}_1)} e^{i{\bf R}_{i_2}({\bf p}_1 - {\bf p}_2)} e^{i{\bf
    R}_{i_{n+1}}({\bf p}_n -{\bf p})} G({\bf p}_1,z)\ldots G({\bf p}_n,z)
\end{multline}\end{widetext}
where $[x]$ is the integer part of $x$, $\mathcal{P}\{i_1,\ldots,i_{n+1}\}$
denotes a permutation of the indices $\{1,2,\ldots,n+1\}$, and the angular
brackets stand for cumulant averages defined from
\begin{equation}
  \label{eq:cumulant}
  \left\langle \exp V\right\rangle_{av}  = \exp \left\{\sum_{n=1}^{\infty}
    \frac 1{n!}\left\langle V^n\right\rangle_c\right\}\ .
\end{equation}
Only the momentum variables and momentum-dependent functions are of
importance. We denote $\mathcal{D}_n$ the sum of all diagrams with $n$
internal fermionic lines with internal momenta
$\mathbf{p}_1,\ldots,\mathbf{p}_n$. Then representation
\eqref{eq:Sigma_expansion} can be simplified to
\begin{widetext}
\begin{equation}
  \label{eq:Sigma_exp_reduced}
   \Sigma({\bf p},z) = \langle
V\rangle_{av} + \sum_{n=1}^\infty \frac 1{N^n} \sum_{{\bf
    p}_1,\ldots,{\bf p}_n} \sum_{{\cal D}_n} X_{{\cal D}_n}(V;{\bf p},{\bf
  p}_1,\ldots, {\bf p}_n) G({\bf p}_1,z)\ldots G({\bf p}_n,z)\ .
\end{equation}
We use notation $ \label{eq:pm_def} G_\pm({\bf p}) = G({\bf p}\pm{\bf
  q}/2,z_\pm)$ and $\Sigma_\pm({\bf p}) = \Sigma({\bf p}\pm{\bf
  q}/2,z_\pm)$ and apply expansion \eqref{eq:Sigma_exp_reduced} for the
self-energy difference. We obtain
\begin{multline}
  \label{eq:Delta_Sigma}
  \Delta \Sigma({\bf p}) = \Sigma_+({\bf p}) - \Sigma_-({\bf p}) =
  \sum_{n=1}^\infty \frac 1{N^n} \sum_{{\bf p}_1,\ldots,{\bf p}_n}
  \sum_{{\cal D}_n} X_{{\cal D}_n}(V;{\bf p},{\bf p}_1, \ldots,
  {\bf p}_n)\\
  \left[G_+({\bf p}_1)\ldots G_+({\bf p}_n) - G_-({\bf p}_1)\ldots G_-({\bf
      p}_n)\right] \ .
\end{multline}
The difference of the products of one-electron propagators can further be
rewritten to a sum
\begin{multline}
  \label{eq:Difference_G}
  G_+({\bf p}_1)\ldots G_+({\bf p}_n) - G_-({\bf p}_1)\ldots G_-({\bf
    p}_n) = \sum_{i=1}^n G_+({\bf p}_1)\ldots G_+({\bf p}_{i-1})\\
  \times \Delta G({\bf p}_i) G_-({\bf p}_{i+1})\ldots G_-({\bf p}_n)
\end{multline}
\end{widetext}
where left (right) to the difference $\Delta G(\mathbf{p}_i)$ only
$G_+(\mathbf{p}_j)$ ($G_-(\mathbf{p}_j)$) appear. We sum all diagrams for
the fixed difference of the one-electron propagators, being now
two-particle irreducible diagrams from the electron-hole channel. We then
come to a new representation
\begin{equation}
  \label{eq:Delta_Sigma_G}
  \Delta \Sigma({\bf p}) = \sum_{n=1}^\infty \sum_{i=1}^n \frac 1{N}
  \sum_{{\bf p}_1} \Lambda^{(i,n-i)}_{{\bf p p}_i}(z_+,z_-;{\bf q}) \Delta
  G({\bf p}_i)
\end{equation}
where $\Lambda^{(n,i)}_{{\bf p p}_i}(z_+,z_-;{\bf q})$ is a sum of
two-particle irreducible diagrams with $n$ internal one-electron lines of
which $i$ lines carry energy $z_+$ and $n-i$ lines energy $z_-$.

Last but very important step in the proof of identity
\eqref{eq:VWW_identity} is an assumption that removing the difference
$\Delta G$ from Eq.~\eqref{eq:Delta_Sigma_G} and fixing the internal
variable $\mathbf{p}_i=\mathbf{p}'$ does not change summability of the
perturbation expansion. If so, we can write
\begin{equation}
\label{eq:Lambda_expansion} \Lambda_{{\bf p p}'}(z_+,z_-;{\bf q}) =
\sum_{n=1}^\infty \sum_{i=1}^n \Lambda^{(i,n-i)}_{{\bf p p}'}(z_+,z_-;{\bf
q})\ .
\end{equation}
The equality holds if and only if the perturbation
expansion in the random potential for the two-particle irreducible vertex
$\Lambda_{{\bf p p}'}(z_+,z_-;{\bf q})$ converges point-wise for the chosen
values of independent variables
$z_+,z_-,\mathbf{p},\mathbf{p}',\mathbf{q}$. This is a rather severe
restriction on applicability of the Ward identity \eqref{eq:VWW_identity}.
It says that the Ward identity is valid only in cases where no selective
rules for sums of two-particle irreducible diagrams apply, i.~e., all
classes of diagrams are equally important.  This happens when the
irreducible two-particle vertex is regular (bounded). However, physically
the most interesting situation occurs when due to backscatterings a Cooper
pole appears in the irreducible vertex $\Lambda$. Thanks to this pole only
a class of relevant (crossed electron-hole) diagrams determines the
low-energy behavior of the two-particle vertex. In the asymptotic region of
the Cooper pole in the electron-hole irreducible vertex, the Ward identity
\eqref{eq:VWW_identity} \textit{cannot be proved}. In fact it gets violated
whenever we take into account a selected series of diagrams dictated by
causality of the approximation or leading to a divergence and we have to
consider a nonperturbative solution for the two-particle irreducible
vertex. Each selected (relevant) diagram of order $n$ from the formal
expansion of the singular two-particle vertex generates in the Ward
identity $n-1$ irrelevant diagrams to complete the sum from the right-hand
side of Eq.~\eqref{eq:Difference_G} so that the self-energy difference can
be represented via the vertex $\Lambda$. The irrelevant diagrams are
neglected in approximations with selected dominant classes of diagrams
violating then identity \eqref{eq:VWW_identity}.  Hence summability of the
perturbation expansion for the two-particle irreducible vertex is an
additive assumption in the proof of the Ward identity
\eqref{eq:VWW_identity}. There is no nonperturbative proof of this identity
except for the homogeneous case, $q=0$, where it is a consequence of the
Velick\'y-Ward identity \eqref{eq:VW_momentum}.

\end{document}